\documentclass{emulateapj}

\newcommand{\gtts}{G337.2$-$0.7}
\newcommand{\xmm}{$XMM$-$Newton$}
\newcommand{\chandra}{{\it Chandra}}
\newcommand{\atca}{{\it ATCA}}
\newcommand{\asca}{{\it ASCA}}
\newcommand{\sibf}{$3.50^{+0.24}_{-0.62}$} %  Si best fit abundance with errors
\newcommand{\sbf}{$5.04^{+0.32}_{-0.82}$}  %  S  ``                          ''
\newcommand{\casill}{$2.65$}  %  Ca/Si 99% lower limit
\newcommand{\cate}{$1.0$} % Ca temperature in 2temp fit  
\newcommand{\site}{$0.75$} % Si temperature in 2temp fit 
\newcommand{\kms}{km~s$^{-1}$}

\begin{document}

\title{Can Ejecta-Dominated Supernova Remnants be Typed from their
  X-ray Spectra? The Case of G337.2$-$0.7}


\author{Cara E. Rakowski\altaffilmark{1}}
\email{crakowski@cfa.harvard.edu}

\author{Carles Badenes\altaffilmark{2}}
%\email{badenes@physics.rutgers.edu}

\author{B. M. Gaensler\altaffilmark{1,}\altaffilmark{3}}
%\email{bgaensler@cfa.harvard.edu}

\author{Joseph D. Gelfand\altaffilmark{1}}
%\email{jgelfand@cfa.harvard.edu}

\author{John P. Hughes\altaffilmark{2}}
%\email{jph@physics.rutgers.edu}

%\and

\author{Patrick O. Slane\altaffilmark{1}}
%\email{pslane@cfa.harvard.edu}


\altaffiltext{1}{Harvard-Smithsonian Center for Astrophysics, 60
  Garden Street Cambridge, MA 02138}
\altaffiltext{2}{Rutgers, the State University of New Jersey, 
  136 Frelinghuysen Road, Piscataway, NJ 08854}
\altaffiltext{3}{Alfred P. Sloan Research Fellow}

\begin{abstract}

In this paper we use recent X-ray and radio observations of the
ejecta-rich Galactic supernova remnant (SNR) G337.2$-$0.7 to determine
properties of the supernova (SN) explosion that formed this source. 
\ion{H}{1} absorption measurements from the 
Australia Telescope Compact Array (ATCA) constrain the distance to
G337.2$-$0.7 to lie between $2.0 \pm 0.5$ and $9.3 \pm 0.3$ kpc. 
Combined with a clear radio image of the outer blast-wave, this
distance allows us to estimate the dynamical age 
(between 750 and 3500 years)
from the global X-ray spectrum obtained with the \xmm\ and \chandra\
observatories. 
The presence of ejecta is confirmed by the pattern of
fitted relative abundances, which show Mg, Ar
and Fe to be less enriched (compared to solar) than Si, S or Ca, and
the ratio of Ca to Si to be 3.4$\pm$0.8 times the solar
value (under the assumption of a single electron temperature and single
ionization timescale).  
With the addition of a solar abundance component for emission
from the blast-wave, these abundances (with the exception of Fe) resemble 
the ejecta of a Type Ia, rather than core-collapse, SN. 
Comparing directly to models of the ejecta and blast-wave X-ray emission
calculated by evolving realistic SN Ia explosions to the remnant stage
allows us to deduce that one-dimensional delayed detonation and pulsed
delayed detonation models can indeed reproduce the major features of
the global spectrum. 
In particular, stratification of the ejecta, with the
Fe shocked most recently, is required to explain the lack of prominent
Fe-K emission.

\end{abstract}




\defcitealias{1995ApJS..101..181W}{WW95}

\section{Introduction}

A supernova (SN) explosion marks the death of a star, whether that be the 
gravitational collapse of a massive star that has run out of nuclear
fuel at its core (core-collapse) or 
the disruption of a white dwarf by runaway nuclear reactions (Type Ia).
The long-lived remnants of SN explosions provide a rich 
tapestry of information about the metal-rich ejecta produced in a SN, as 
well as the environment into which the remnant is expanding. 
X-ray spectra of supernova remnants (SNRs) are particularly useful
given the extremely high shock temperatures and the fact that most
cosmically abundant elements produced in the explosion have diagnostic
K-shell lines in the 0.5 to 8.0 keV X-ray bandpass.   
Furthermore, the expansion of the ejecta allows their structure to be
spatially resolved.  Because X-ray emission from the ejecta can be
identified thousands of years after the explosion, 
there exists a sample of young nearby SNRs despite a galactic 
SN explosion rate of only $\sim$1 per century. 
Given the diversity within the classes of core-collapse and Type Ia
explosions, and the variety of open issues regarding their
explosion mechanisms, it behooves us to study all the young
SNRs in our Galaxy and nearby in the hope of gaining an understanding
of the supernova explosions that originated them.  

Evidence for metal-rich ejecta in %\object[SNR 337.2-00.7]{
SNR G337.2$-$0.7  was first found based 
on the strength of the emission lines in the \asca\ X-ray spectrum 
(Rakowski et al. 2001). In that analysis, Si and S were confidently 
shown to be over-abundant compared to 
their solar abundances, and the pattern of metal abundances relative to 
each other was also highly non-solar. However the abundance ratios 
between the metals were not sufficiently well constrained to 
determine the origin of the SNR. In this paper, we analyze 
new observations of \gtts\ with \chandra\ and \xmm , and we 
complement these X-ray observations with radio data from the Australia 
Telescope Compact Array (ATCA). 
In our analysis, we devote special 
attention to the limitations of the techniques that are commonly used 
to measure elemental abundances, 
and how they influence our ultimate goal of 
determining the origin of this supernova remnant and contributing to our 
current understanding of SN explosions. 
Great care must be taken when interpreting X-ray
spectra from SNRs because this emission will be dominated by the
hottest, densest highly ionized material. Abundant elements
can be hidden if they are in a low density low temperature region,
not co-mingled with the bulk of the ejecta. Even amongst the bright
ejecta, if the standard single electron temperature, single ionization
timescale models are used, hot plasma may be misinterpreted as 
overabundances.  

From the theoretical side, numerical simulations of SNe are rapidly
increasing in sophistication.  The inclusion of magnetic fields, angular
momentum, asymmetries, and instabilities in the flow all require a 3-D
treatment and are beginning to be explored in both core-collapse
and Type Ia explosions. 
In this introduction we highlight some of the core issues under debate
in the theory of SN explosions, which will set the context for our
in-depth analysis of \gtts .

A key question for core-collapse explosions is what rejuvenates
the shock after core-bounce. 
When the in-falling material forces the core of the star to
    such a high density that a neutron-rich nuclear composition is
    favored, the core stiffens and pushes back (``core-bounce'').
However, in spherically symmetric simulations, this
shock will stall against the force of the continually
infalling material and fail to cause an explosion.
A multi-dimensional mechanism is required to revitalize the shock,
with the contenders including jet-induced explosions 
\citep{1999ApJ...524L.107K,2002ApJ...568..807W,2002ApJ...565..405M,2003ApJ...584..954A},
neutrino driven convection 
\citep{2000ApJ...541.1033F,2003A&A...408..621K,2003PhRvD..68d4023K,2005ASPC..332..358B},
and turbulent dissipation of magneto-rotational energy  
\citep{2005ApJ...620..861T}.
Magneto-hydrodynamically (MHD) driven jets have particularly been
proposed as a mechanism for gamma-ray burst (GRB) explosions 
\citep{2002ApJ...568..807W,2002ApJ...565..405M}.
The magnetic field may
be pre-existing or it could be amplified by the magneto-rotational
instability \citep{2003ApJ...584..954A}. 
There have also been improvements in neutrino transport calculations
\citep{2004ApJ...609..277L} and the incorporation of rotation into
neutrino-driven convection models 
\citep{2000ApJ...541.1033F,2003PhRvD..68d4023K,2005ASPC..332..358B}.
Including rotation naturally leads
to a bi-polar structure in the neutrino flux and accretion of mass,
such that the increased neutrino heating is sufficient and directed
enough to revitalize the shock near the poles and explode the
star. Alternatively, the rotational energy of the in-falling material
itself can be converted into another source of heating behind the
stalled shock by turbulent dissipation of the differential shear via
the magneto-rotational instability \citep{2005ApJ...620..861T}.

Simulations of white dwarf explosions are also progressing. 
The main difficulty is in constructing a realistic explosion
that reproduces the abundances and stratification
of the ejecta inferred from Type Ia spectra
\citep[e.g.][and references therein]{2005PASP..117..545B}. In one dimension,
supersonic flame propagation (detonation) can burn almost the entire
star to Fe-group elements because the star retains the original
density ahead of the flame.
A purely sub-sonic flame (deflagration) will push the 
star outwards, reducing the density of the material ahead of the
shock. This causes the flame to be quenched earlier than in the
detonation case and leave behind a large outer shell of unburnt
material.\footnote{This early quenching of the flame in one-dimensional
  deflagrations is characteristic of realistic flame velocities that have
  superceded the ``fast deflagration'' model W7 
  of \citet{1984ApJ...286..644N}.}  
Neither pure detonation nor pure 
deflagration produces the necessary quantities of intermediate mass
elements (between O and Fe). The {\it ad hoc} solution in
one dimensional simulations that successfully matches observation has been to 
invoke a transition from sub-sonic to supersonic flame propagation at
some critical density, a ``delayed detonation''
\citep{1991A&A...245..114K}.
Phenomenologically,
this allows the star to expand first before the
detonation front runs through it, thus creating a wider region 
of incomplete burning where Si, S, Ar, and Ca are produced before
the flame is quenched. Considerable effort is going into simulating the actual
turbulent burning front over a range of different scales to look for
evidence of such a transition \citep{2003ApJ...588..952R,2004A&A...420L...1R,
2005A&A...429L..29R,2004ApJ...606.1029B,2004ApJ...608..883B}.
However, although the results of multidimensional physical effects and
micro-physics of the flame can be worked into the one-dimensional
models, the flame propagation is inherently three-dimensional. 
Current three dimensional simulations have generally refrained from
artificially inducing a transition to supersonic burning, but
these pure deflagrations still suffer from a deficit of
intermediate mass products just as they did in one dimension. The
other difficulty is that the three dimensional simulations are almost
all well-mixed, unlike the stratification seen in Type Ia spectra
\citep{2005A&A...437..983K}. 
\citet{2004ApJ...612L..37P} have produced preliminary 
models that retain a stratified elemental structure in three dimensions, 
by initiating the burning at a single, slightly off-center point and 
having the resulting hot bubble break out on the WD surface to trigger a 
'gravitationally confined detonation'.
However if the seeds for instabilities are present at the
earliest times they will rapidly run away to thoroughly mix burnt and
unburnt material \citep{2005A&A...429L..29R}.
  
%(e.g. Ropke et al 2004, Bell et al 2004). 
%Plewa, Calder and Lamb (2004)(e.g. Ropke \& Hillebrandt 2005)

\begin{figure*}
\includegraphics[width=6.4in]{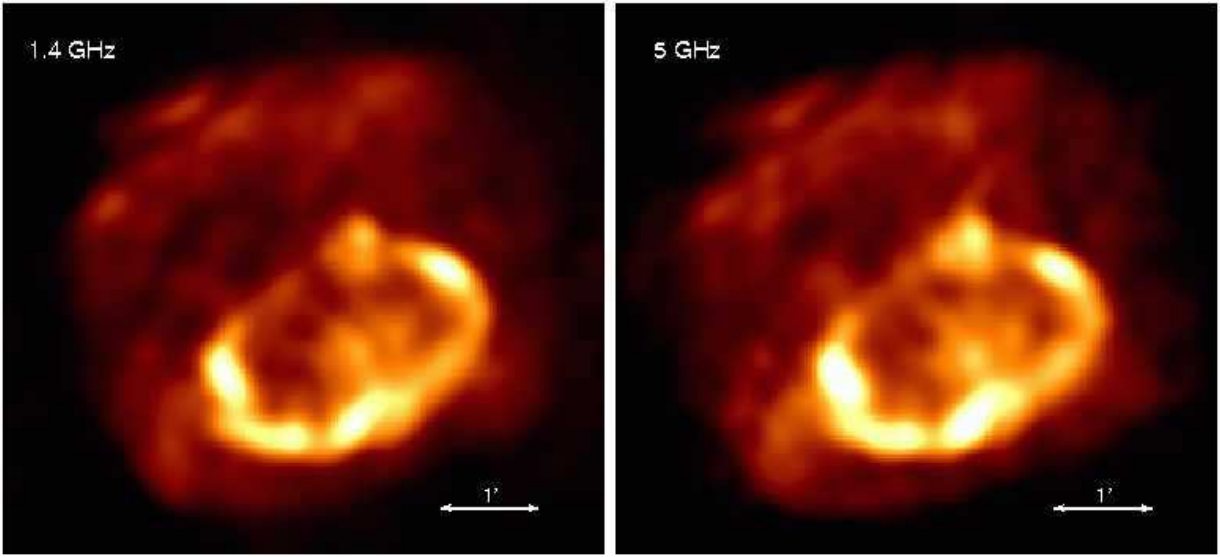} 
\figcaption{\atca\ 1.4GHz (left) and 5GHz (right) images of SNR \gtts\
  smoothed to 15\arcsec resolution (North:up East:left). 
  The intensities are linearly
  scaled from 0 to 20 mJy per beam (1.4GHz) and 0 to 10 mJy per beam (5GHz).
  \label{radio_5GHz}}
\end{figure*}

The study of young SNRs is playing an increasingly
important role in uncovering the SN explosion mechanism. For example,
the ``jet'' in the northeast corner of Cassiopeia A (Cas A), first detected
in the optical fast-moving S-rich knots by \citet{1970ApJ...162..485V},
has been cited as a possible example of the jet-induced
core-collapse explosion model \citep{1999ApJ...524L.107K}. 
However, \citet{2000ApJ...528L.109H} found that the composition of the
X-ray emitting 
portions of the jet was more easily explained by Si-rich explosive Oxygen
burning with little or no Fe, suggesting that the jet material
originated further out in the exploding star. In contrast, to the
southeast, \citet{2000ApJ...528L.109H} found knots of almost pure Fe out past
the bulk of the Si-rich ejecta. In deeper \chandra\ observations,
\citet{2003ApJ...597..362H} confirmed the composition of these knots,
finding others that could be explained by a pure Fe plasma. The
high-velocity Fe-rich bullets in Cas A are reminiscent of the bullets
of near-core material that would be ejected as a convection cycle
broke through the stalled shock in the neutrino-driven explosion
models of  \citet{1995ApJ...450..830B}. This model was motivated by  
the early detection of Fe in the spectrum of SN 1987A. 
However, reproducing both features, Ne-rich bullets and a ``jet'' that
originated further from the core, in one model is somewhat
problematic \citep{2005ASPC..332..358B}. 
 
Type Ia SNRs have also shown interesting structures that lead to
constraints on the explosion mechanism. 
Specifically, the stratification observed in Type Ia SN spectra 
is also seen in Type Ia SNRs. In the remnant of SN 1006 multiple lines
of evidence point to stratification of the Si and Fe ejecta both from
ultra-violet absorption lines of Si and Fe \citep[e.g.][]{1997ApJ...481..838H}
and the lack of Fe-L emission in the X-rays \citep{2003ApJ...587L..31V}.
In Tycho's SNR,
spatially resolved spectroscopy of the ejecta supports 
the idea of stratification of conditions and composition
\citep{1995ApJ...441..680V,1997ApJ...475..665H,2003ApJ...593..358B}. 
Detailed comparisons of Tycho to one and three dimensional 
Type Ia SN explosion models have shown that one-dimensional,
stratified ``delayed detonation'' models best reproduce the 
spatial composition of Tycho's SNR while recent well-mixed
three-dimensional deflagration models fail
\citep{2005astro.ph.11140B}.
%\citep{2005ApJ...624..198B}.

These considerations regarding SN explosions 
are the context for our interpretation of the new \xmm , \chandra\ and 
ATCA X-ray and radio observations of \gtts , an ejecta-rich remnant 
that still bears a clear imprint from the 
explosion that produced it.
In Section~2 we outline the radio and X-ray observations and
data reduction.
The radio and X-ray morphologies of \gtts\ are compared in Section~3 
and an estimate of the distance to the remnant is found from
\ion{H}{1} absorption measurements in Section~4. 
Section~5 discusses the results of non-equilibrium 
ionization (NEI) modeling of the global X-ray spectrum of the remnant. The 
validity of utilizing a single temperature model for the global
spectrum is investigated with respect to  spatial variations
throughout the remnant in Section~6. We then  
proceed to compare the results of our global NEI model to the predicted 
abundances from core-collapse and Type Ia SN models in Section~7.
A detailed comparison of the global spectrum to the Type Ia model SNR 
spectra from \citet{2003ApJ...593..358B} is performed in Section~7.1. 
We conclude by incorporating the results of these various
investigations into a single picture of the ejecta-rich SNR \gtts\ 
(Section 8).

\section{Data, Processing, and Calibration \label{redux}}

\subsection{Radio Observations}

We have obtained radio observations of \gtts\ at multiple frequencies
using the ATCA, a six-element aperture
synthesis telescope located near Narrabri, Australia \citep{fbw92}.
Observations centered at frequencies of 4672, 4800 and 5312~MHz were
carried out on 1999~July~19/20 in the 0.75D configuration, using
a bandwidth of 128~MHz at each frequency and for a total effective
integration time of 6.5~hours. Subsequent observations were carried out
on 2004~Jul~13 (in the 6A configuration) and 2004~Dec~18 (in the 1.5D
configuration), each recording data simultaneously at 1384~MHz (using
a 128~MHz bandwidth) and in the \ion{H}{1} line (using a 4~MHz bandwidth)
for a total integration time of 20.5~hours.  To recover the full
range of spatial scales in the \ion{H}{1} line, we also made use of
shorter interferometer spacings from the Southern Galactic Plane
Survey (SGPS) \citep{2005ApJS..158..178M}\footnote{\url{http://www.atnf.csiro.au/research/HI/sgps/queryForm.html}}. 

The ATCA data were edited and calibrated using standard techniques.
Observations at 4672, 4800 and 5312~MHz were then combined into a single
image (hereafter referred to as the 5-GHz image), while the 1384~MHz
data were imaged separately (hereafter referred to as the 1.4-GHz image).
Both images were deconvolved using maximum entropy, smoothed
with a Gaussian restoring beam, and then corrected for primary beam
attenuation. Residual atmospheric gain variations were significant for
the 5~GHz data; these were corrected using two iterations of phase
self-calibration. 

The final 1.4-GHz image has a resolution of $8''\times6''$
and a sensitivity of 70~$\mu$Jy~beam$^{-1}$, while the 5-GHz
image has a resolution of $12''\times10''$ and a sensitivity
of 90~$\mu$Jy~beam$^{-1}$. The resulting maps are shown in
Figure \ref{radio_5GHz}, 
and in both cases are expected to be sensitive
to all spatial scales associated with the SNR. The corresponding flux
densities of the SNR are $1.55\pm0.05$~Jy at 1.4~GHz and $0.93\pm0.02$~Jy
at 5~GHz, in both cases using observations of PKS~B1934--638 referenced
to the scale of \cite{rey94}. This implies a spectral index $\alpha =
-0.4$ ($S_{\nu} \propto \nu^{\alpha}$), which is well within the
typical range for SNRs. 

For the \ion{H}{1} observation, continuum emission was subtracted from the
data in the visibility plane. The data were then imaged at a velocity
resolution of 3.3~\kms, using a spatial filter aimed at excluding \ion{H}{1}
emission on larger scales while retaining sensitivity to absorption
on spatial scales corresponding to structure in the SNR. The emission cube 
was then deconvolved using the CLEAN algorithm, and restored to a
resolution of $39''\times36''$. 

\begin{figure*}
\includegraphics[width=6.4in]{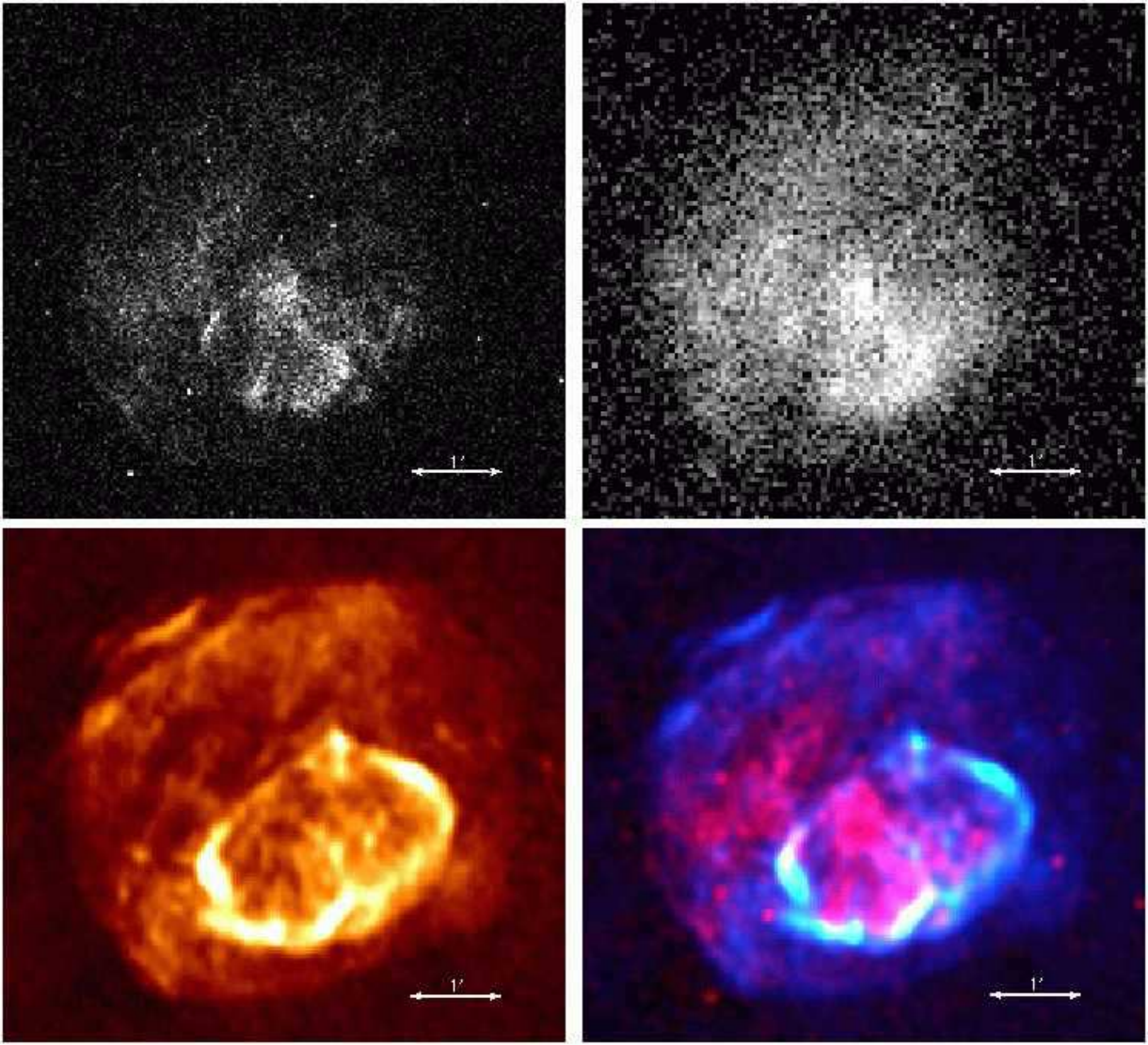}
\figcaption{X-ray and radio images of SNR \gtts . Upper left:
  broadband 1 to 5 keV \chandra\ image, binned by four
  pixels, linearly scaled from 1 to 12 counts per bin.
  \label{chandra_bw} 
  Upper right: combined \xmm\ MOS1 and MOS2 image of \gtts\ from 1 to 5
  keV. Square-root scaling was chosen in this image to emphasize the
  fainter diffuse emission, almost filling the extent of the radio
  remnant. \label{mos12_bw}
  Lower left: Full resolution 1.4 GHz \atca\ image linearly scaled
  from 0 to 4 mJy per beam. \label{radio_14GHz}
  Lower right: Overlay of the broadband \chandra\ X-ray
  (red) and the 1.4 GHz \atca\ radio (blue-green) images. An arbitrary
  scaling was chosen to show both faint (blue) and bright (green)
  radio emission against a smoothed \chandra\ X-ray image (Gaussian
  kernel width 3 four-pixel bins) shown in red. \label{radio_xray} 
}
\end{figure*}

\subsection{\xmm\ and \chandra\ X-ray Observations}

\gtts\ was observed for 38.4 ks by \xmm\  on 2001 Feb 18,
%\dataset[ADS/Sa.XMM#obs/0087940101]{
observation ID 0087940101. 
Standard reduction
and processing of the three EPIC datasets were carried out using
available XMM-SAS 6.0.0 software. Starting from the level 1 event files, 
the latest calibrations were applied with the ``emchain'' and ``epchain'' 
tasks. The events were then filtered to retain only the ``patterns'' and 
photon energies likely for X-ray events: 
patterns 0 to 4 (single and double pixel events) and 
energies 0.2 to 15.0 keV for the PN,
patterns 0 to 12 (including triple and quadruple events) and 
energies 0.2 to 12.0 keV for the two MOS instruments. 
Events near hot pixels, outside the field of view or otherwise suspect 
are ``flagged'' within the event file and were eliminated using the most 
conservative screening. 
Within the good time intervals, no significant flares
were detected, and the background count-rate was consistent with that
measured in average quiescent times. 
The effect of any undetected soft flares was
minimized by the use of local background from the same
observation that was extracted from an ellipsoidal annulus outside 
the supernova remnant. Spectra extracted from the PN, MOS1 and
MOS2 cameras were 
fitted simultaneously to maximize statistical significance
(after consistency between instruments was verified). 
Detector response files and effective areas for the extracted spectra
were also generated with the standard XMM-SAS 6.0.0 software. 

\chandra\ observed  \gtts\  for 48.8 ks with the back-side illuminated CCD 
ACIS S-3, beginning on 20 August 2002, 
%\dataset[ADS/Sa.CXO#obs/2763]{
observation ID 2763. 
These data were reprocessed using CIAO 3.1 and 
caldb 2.26. The events were first 
calibrated with the appropriate instrument response files using 
``acis\_process\_events,'' then filtered on grade, on energy, 
and to eliminate bad pixels and periods of high background. 
The time-dependent gain correction was applied using the contributed software 
``TGAIN'' from Alexei Vikhlinin (this is the same correction that is now
available as part of ``acis\_process\_events'' in CIAO 3.2). 
All spectra were extracted with their own weighted response files using 
the ``acisspec'' task.
Background contamination, primarily from the Galactic Plane region,
was important for \gtts . For simplicity, an annulus around the SNR was
used to subtract local background directly from each spectra (weighted
by the ratio of the areas of the background and source regions).
Images of \gtts\ from \xmm\ and \chandra\ are shown in Figures
\ref{chandra_bw} and \ref{3colorbox}.

\section{Radio and X-ray Morphology \label{morph}}

The 1.4 GHz and 5 GHz images of \gtts\ are very similar when
convolved to the same spatial scale (Figure \ref{radio_5GHz}).
Both show faint
diffuse emission from a 4.5\arcmin\ $\times$ 5.5\arcmin\ (diameter)
SNR.
A wider radio image of the Galactic plane \citep{1999ApJS..122..207G}
shows \gtts\ to be in an isolated region off the plane.
While not a diagnostic of the progenitor type, one would expect a
larger number of
SNRs generated by 
Type Ia SNe than by core collapse SNe in this region, because Type Ia 
progenitor systems have more time to drift away from their birthplaces
at low latitudes before the explosion than more massive stars do.
The sharp edges that presumably demarcate the outer blast-wave
appear to be somewhat squared off and only limb-brightened in a few
regions.  It is important to note that this is a genuine feature rather
than an artifact due to the use of ``clean boxes'' --- i.e., while the
deconvolution process was constrained by the use of region selection,
in both cases the boundaries of the regions used lay well outside the
sharp edges seen here.

The most prominent feature at both radio frequencies 
is a bright elliptical ring 
2.0\arcmin\ $\times$ 3.2\arcmin\ in extent in the south part of the SNR.
Such a brightening in the radio must indicate an enhancement
in the magnetic field, electron density or both.
The ring is incomplete and clumpy, appearing primarily as short bright arcs 
with one spoke into the center and a bright clump in the north 
that extends outside the main ellipse. 
However, with the possible exception of the bright northern clump that
may have a slightly flatter spectral index ($\sim - 0.25$),  the ring
does appear to be a coherent structure with a constant spectral index,
$-0.5\pm0.1$ and weak linear polarization at 5~GHz (up to peaks of 10\%
fractional polarization at the ends of the major-axis).

The \chandra\ and \xmm\ (MOS1$+$MOS2) images are shown in the top
panel of Figure \ref{chandra_bw}. 
The higher resolution \chandra\ image has been
scaled linearly to emphasize the bright clumps of X-ray emission near
the radio ring, while the deeper \xmm\ image is square-root-scaled to
show the faint diffuse emission that extends throughout most of the
radio shell. 
The bottom panel of Figure \ref{chandra_bw} shows the 1.4 GHz image at full
resolution on the left and a 3-color overlay of the \chandra\ X-ray
(red) and \atca\ 1.4~GHz (blue$+$green) images on the right. 
In this overlay, the fainter radio emission is shown in blue while the bright
radio ring has been emphasized by displaying it in green
(turquoise). Purple and white regions indicate an overlap between the
X-ray emission and the faint or bright radio emission, respectively.

Looking outside of the bright radio ring,
a wide band of X-ray emission north-east of the ring lies where 
the radio emission is weakest. A faint radio
filament traces the brightest portion of this swath of X-ray
emission (compare the 1.4~GHz image with the 3-color overlay). 
Elsewhere some amount of faint X-ray emission accompanies most of
the faint radio emission inside the rim, with the possible exceptions
of the narrow region south of the bright ring and the western-most side
of the remnant. Along the limb-brightened portions of the radio rim, 
there are suggestions of X-ray emission above the
background with one clear identification of an X-ray filament aligned
with the radio rim in the south-east. This filament is brightest
in the X-rays where it is faint in the radio, but they are clearly
related since they share the same small indentation (located at the
brightest part of the X-ray emission of the filament).

As for the bright ring itself, there is one short arc of X-ray bright
material aligned with the northeastern tip of the radio
ellipse, beginning just as the radio surface brightness is
dropping. In the southern portion of the ring, the arcs of bright
radio emission extend further out than their X-ray counterparts, while
in the north-west no X-ray emission is seen from the strong radio arcs. 
The radio clump in the north of the ring shows only faint X-ray
emission.
Vice-versa the X-ray bright clump on the north-eastern side aligns
with a particularly faint spot in the 1.4 GHz image, and 
the spoke of X-ray emission across the minor axis of the ellipse
does not align well with the radio protrusion from the south rim.

The meaning of the differences and correspondences between the X-ray
and radio morphologies is not obvious, but clearly deserves more attention.
What insights into the origin of the bright radio ring could the X-ray
spectral observations divulge?
If the ring demarcates the current position of the reverse shock, 
that would imply higher abundances in the X-ray spectrum from that
region. An interaction with  some structure in the
circumstellar or interstellar medium (ISM) should be reflected in lower
abundances of high-Z elements. 
Some feature intrinsic to the explosion itself would be likely to have
different abundance ratios of heavy elements than elsewhere. 
We address these issues further in Sections~6 and~7.1.

\begin{figure}
\noindent \includegraphics[width=3.3in]{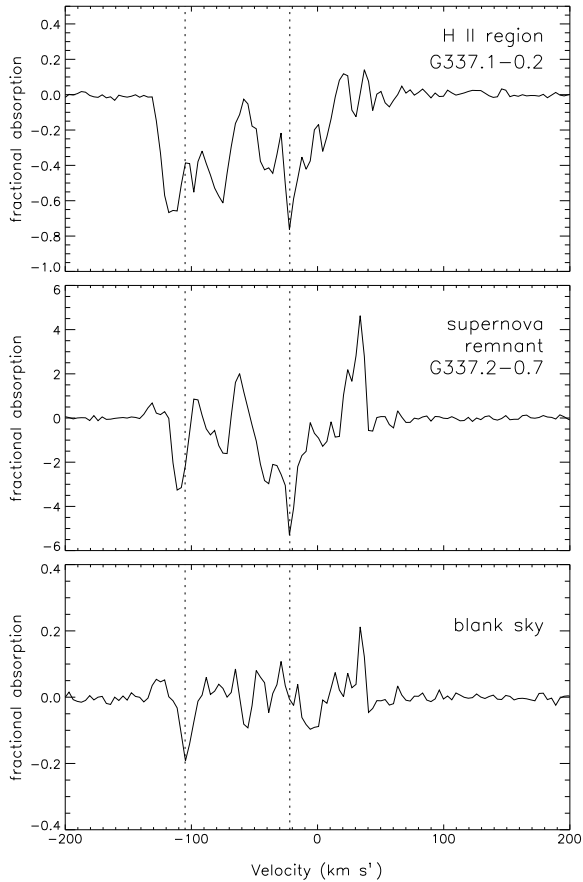} \\
\figcaption{\ion{H}{1} absorption spectra from three regions in the
  field, G337.1$-$0.2 an unresolved \ion{H}{2} region $\sim$11~kpc
  distant, the supernova remnant itself, and a blank-sky
  position about 2\arcmin\ southwest at RA 16:39:14.151, Dec
  -47:53:43.52 (J2000). 
  All absorption
  features to the \ion{H}{2} region are point-like and hence real. 
  Only the $-22$\kms\ feature in the remnant spectrum
  distinctly followed the morphology of the SNR. Absorption near 
  $-113$\kms\ extended in a wide swath a few arcminutes in size. The
  bottom panel shows a sample spectrum from a source free region
  2\arcmin\ southwest of the remnant exhibiting clear
  absorption at $-110$\kms . Fractional absorptions relative to the
  1.4 GHz continuum are plotted. Note that for the SNR these exceed an
  absolute value of 1 indicating the presence of \ion{H}{1}
  self-absorption over this region, such as might be
  expected from the large molecular cloud seen in $^{12}$CO.
   \label{HIspec} }
\end{figure}

\section{Constraints on the Distance to SNR \gtts \label{distance}}

To constrain the distance to SNR~\gtts\ via \ion{H}{1} absorption, we
examined both the ATCA \ion{H}{1} cube centered on \gtts, plus SGPS
\ion{H}{1} data on 
surrounding fields, both on a channel-by-channel basis. Real
absorption features to \gtts\ should be significantly deep and conform
to the shape of the remnant. Figure \ref{HIspec} plots the fractional
absorption for three targets in the field, 
G337.1$-$0.2 (an unresolved bright \ion{H}{2} region $\sim30^{\prime}$
northwest of the SNR, at a distance of 11~kpc \citet{ccdd99}), 
the supernova remnant itself and a blank-sky
position about 2\arcmin\ southwest of the remnant.
The strongest absorption features toward G337.1$-$0.2 are at Local
Standard of Rest (LSR) velocities of $-$22~\kms\ and $-$116~\kms\
\citep[see also Fig.~4 of][]{sggf97}. The channel images show both to
be entirely point-like and thus truly along the line of sight to this
unresolved 
\ion{H}{2} region. SNR \gtts\ shows absorption at $-$22~\kms\ which
conforms to the shape of the remnant, but the next strongest
absorption feature at $\sim -$116~\kms\ extends evenly across a wide
swath of 
the image as can be seen in the blank-sky spectrum in the bottom
panel of Figure \ref{HIspec}. Given that the $^{12}$CO survey of
\citet{2001ApJ...547..792D} shows that cold absorbing gas at this
velocity covers the sight-lines toward both \gtts\ and G337.1$-$0.2 we
take the lack of absorption specific to the SNR at $-$116~\kms\ to be 
significant and to indicate that the SNR is closer than this cloud.

We thus adopt $-$22~\kms\ and $-$116~\kms\ as upper and lower limits,
respectively, on the systemic velocity of \gtts. Using the Galactic
rotation curve of \cite{1989ApJ...342..272F}, and assuming a distance to
the Galactic Center of 8.5~kpc, we constrain the distance to \gtts\ to be
between $2.0\pm0.5$~kpc and $9.3\pm0.3$~kpc, where we have assumed typical
uncertainties of $\pm7$~\kms\ in the velocity of each \ion{H}{1} feature.
In the final analysis we also consider the possibility that general
absorption in the region around \gtts\ has obscured a true absorption
feature at $-116$\kms\ , i.e. that we should ignore the upper limit 
on the distance.

\section{X-ray Spectral Analysis: 
The Global Spectrum of the Remnant\label{nei}}

\subsection{An Aside on the Usage of the Word Abundance \label{aside}}

Given that the primary subject of this paper is to reconcile 
the abundance pattern that we see in the spectra of SNR \gtts\ 
with its radio and X-ray morphology, as well as 
directly with SN explosion model predictions, 
it is worthwhile to first clarify exactly to what we are referring.
When speaking of the abundances there are many things to
consider. The first distinction to make is between ``fitted abundances,''
i.e. how much of any given element is inferred from a spectrum, and
the overall metal production predicted by a SN model.
The fitted abundances generally assume a single electron temperature and
timescale and are actually proportional to
the ``volume emission measure'' (EM)
which is the product of electron density times the ion density
integrated over the emitting the volume. 
Not only might some of the ejecta be at lower temperatures than others
but if the density in a particular layer is too low it simply
will not be seen. 
For example, in the Type Ia SNR models of \citet{2003ApJ...593..358B} 
there are cases where the relative emission measures of elements 
compared to relative masses produced can be different by 3 or 4 
\textit{orders of magnitude}. In fact, the less abundant element
can appear to be the more abundant one 
\citep[see O versus Fe in Figure 4b of][]{2003ApJ...593..358B}. 
Under these circumstances, it is clear that comparing globally fitted 
abundances to the bulk chemical composition obtained from a SN explosion 
model is not always justified, and can lead to erroneous conclusions.
The second hurdle is the notation used for fitted abundances.
Conventionally in NEI modeling,
especially in XSPEC, the ``abundance'' of an element is defined by
a linear factor relative to its ``solar abundance.'' The solar
abundance set is written on the logarithmic ``dex'' scale for the
number of atoms ($n_{X}$) of each element ($X$) relative to the 
number of hydrogen atoms ($n_{\rm H}$) as follows:
$ dex(X) =  12.0 + log(n_{X}/n_{\rm H})$.
In contrast, SN modelers generally report the solar masses
of each element produced, not the number. Once the presence of
ejecta in the emitting region has been detected by the fitted
abundances, it is often the case that the abundances of the metals
relative to each other are better constrained than the number of
atoms relative to hydrogen.\footnote{The line strengths are measured
  relative to the bremsstrahlung continuum which is generated by free
  electrons that are contributed by all elements in the plasma} 
Thus,
throughout this paper we will refer to both the ``abundance'' of
any given element (relative to hydrogen relative to solar) and the
``relative abundances'' or ``abundance ratios'' of one metal to
another. These relative abundances will be given with respect to
their solar ratio. 
For example, a Ca to Si ratio of 2 times solar, does
not imply that there are twice as many Ca atoms as Si atoms but
rather that there are twice as many Ca atoms per Si atom as would
be present in a ``solar abundance plasma.'' 
Further, we will refer to
the set of metal abundance ratios as the ``abundance pattern'' (listed
relative to Si). Normalizing elemental abundances to a given 
``solar'' composition is useful for studies of the ISM, 
but it is often an 
awkward convention for SN ejecta, especially in the case of Type Ia SNe, 
which have no H in the ejecta.

\begin{figure}
\noindent \includegraphics[width=3.4in]{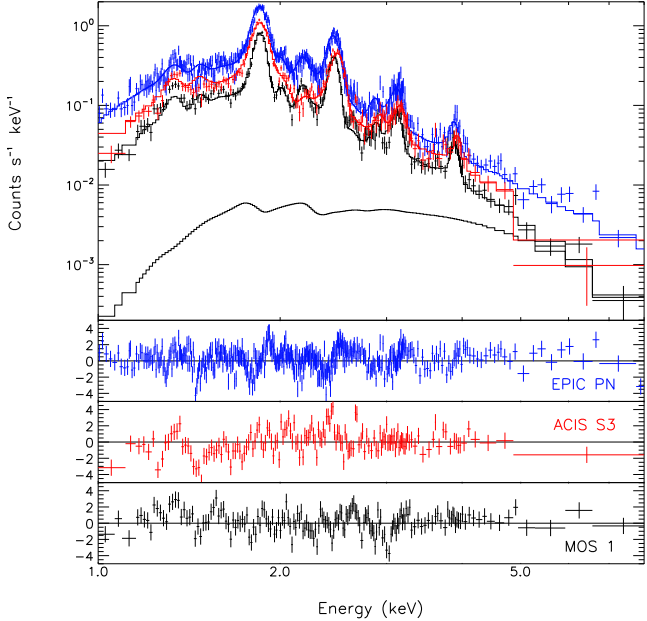} \\
\figcaption{\xmm\ EPIC PN, MOS 1 and \chandra\ ACIS S-3 spectra of the
  overall SNR \gtts . (The MOS 2 spectrum is similar to MOS 1). The
  model is a non-equilibrium ionization thermal plasma model with the
  abundances of Ne, Mg, Si, S, Ar, Ca and Fe allowed to vary along
  with the absorbing column density, the ionization timescale and the
  electron temperature. An additional power-law component
  significantly improved the fit. An example of the contribution of the second
  component to the overall model is plotted here for MOS1.
  The spectra have been re-binned to
  signal-to-noise of 4 for plotting purposes only. \label{neifig}}
\end{figure}

\subsection{NEI Model Fitting Procedure}

The global SNR spectrum contained 
13900, 14500, 23900 and 34200 counts from MOS1, MOS2, PN, 
and ACIS-S3 respectively.
To begin, the spectrum from each instrument was fit
independently with a planar shock NEI model. The shock model used 
is described in \citet{2000ApJ...543L..61H} and allows a variable electron 
temperature, $kT_{\rm e}$, final ionization timescale, $n_{\rm e}t$,
and normalization.\footnote{The model was incorporated for
  our own use as a ``local model'' into XSPEC version 11.3.1 for
  spectral fitting  
\citep[http://xspec.gsfc.nasa.gov/]{1996ASPC..101...17A}. }
This model includes lines from
\cite{1985A&AS...62..197M} that are 
important during the transient ionizing stage as described in
\cite{1994ApJ...422..126H}. 
Planar shock models are widely used as a first approximation to the
spectrum of a complex SNR in order to obtain 
estimates of the bulk properties and abundances of the ejecta (or blast-wave)
by initially assuming that all emitting material is under the same
conditions. 
During the fitting procedure, 
the abundances of all elements were initially held
at their solar ratios, but freed as necessary where the data
could clearly constrain the individual abundances. The NEI model was
convolved with an absorption model (using solar abundances) for the
interstellar column density, $N_{\rm H}$. Where the statistics
warranted it, a power-law component was also included
to account for additional high-energy continuum emission that could be
coming from either non-thermal X-ray synchrotron emission (appropriate
for the power-law model used here) or a second, higher temperature
thermal component (explored further in Sections~7 and~7.1).
Although small shifts in the line centroids between instruments could
be seen \citep[of order 15 keV, similar to those found in \xmm\ analysis
of Kelper's SNR][]{2004A&A...414..545C}
no significant differences in the fits between instruments were found 
for spectra extracted from any region. All small differences in the 
fitted parameters could be explained by the different quantum efficiencies 
for the PN, the MOS, and ACIS-S3 which effected what portion of the 
spectrum had the greatest statistical weight. 

The PN, MOS1, MOS2 and ACIS-S3 spectra of the overall SNR were then fitted
simultaneously with a single NEI model. The only parameter that was
allowed to differ between the instruments was the normalization which
varied slightly, most probably owing to imperfect area calculations
around the chip gaps, hot and dead pixels.  An iterative process of
fitting each of the instruments individually and then in combination was
used for speed to locate the area of parameter space where the best
fit would lie. This was followed by a thorough 
search of the surrounding area of parameter space using all four
instruments to identify the best-fit and constrain the 1$\sigma$
uncertainty ranges of all the parameters. The results of these fits
are given in Table \ref{globalnei}.

\begin{deluxetable}{lcc}
\tablewidth{0pt}
\tablecaption{NEI Parameters for the Global Spectra of 
SNR \gtts \label{globalnei}}
%\tabletypesize{\scriptsize}
\tablehead{
 & \asca \tablenotemark{a} & \xmm\ \& \chandra\ }
\startdata
$N_{\rm H}$ ($10^{22}$ cm$^{-2}$) & $3.5\pm 0.3$ & $3.2\pm 0.1$ \\
$kT_{\rm e}$ (keV) & $0.85^{+0.04}_{-0.03}$ & $0.74^{+0.03}_{-0.01}$ \\
log($n_{\rm e}t$) log(cm$^{-3}$s) & $12.25^{+\inf}_{-0.43}$ 
& $12.02^{+0.05}_{-0.08}$ \\
$n_{\rm e}n_{\rm H}V / 4\pi D^{2}$ cm$^{-5}$ & $1.7\times 10^{12}$ 
& $(2.0\pm 0.4) \times 10^{12}$ \tablenotemark{b} \\
Ne\tablenotemark{c} & $0.0^{+2.1}$  & $4.16^{+0.84}_{-1.64}$ \\
Mg & $0.0^{+2.1}$  & $1.74^{+0.20}_{-0.40}$ \\
Si & $5.4^{+12.2}_{-2.2}$ & $3.50^{+0.24}_{-0.62}$ \\
S  & $4.7^{+9.9}_{-1.8}$ & $5.04^{+0.32}_{-0.82}$ \\
Ar & $2.7^{+4.6}_{-1.1}$ & $1.69^{+0.11}_{-0.26}$ \\
Ca & $2.8^{+7.1}_{-2.8}$ & $11.77^{+0.83}_{-2.04}$  \\
Fe & $6^{+19}_{-3}$ & $0.98^{+0.18}_{-0.29}$ \\ 
Ni &  & $0.0^{+0.126}$ \\
$\Gamma$\tablenotemark{d} &  & $2.2^{+0.2}_{-0.5}$ \\
photons cm$^{-2}$ keV$^{-1}$ s$^{-1}$\tablenotemark{e} &
   & $(4.0 \pm 2.5)\times 10^{-4}$ \\
$\chi$ (dof) & 103.1(67) & 2023(1348) \\ 
$\chi_{r}^{2}$ & 1.539 & 1.501 \\ \hline
\enddata
\tablenotetext{a}{Rakowski et al 2001}
\tablenotetext{b}{Average normalization across all four
  instruments. }
\tablenotetext{c}{Abundances relative to their solar values
  \citep{1989GeCoA..53..197A}}
\tablenotetext{d}{Spectral index of an additional power law component}
\tablenotetext{e}{Normalization at 1 keV}

\end{deluxetable}

\subsection{Results from the Global SNR Spectrum \label{OneNEI}}

First let us consider how the overall SNR spectrum compares with our
previous \asca\  results and conclusions about the remnant's dynamical
state \citep{2001ApJ...548..258R}. 
The overall best-fit column density, electron temperature, ionization
timescale and normalization derived from the \asca\ spectrum are all
reasonably consistent with the more well-constrained \xmm\ and
\chandra\  results.  
The main result from the \asca\  spectrum was the presence of a
significant amount of ejecta as
exhibited by the prominent Si, S and Ar lines. The overabundance of
these three species is clearly demonstrated in the new observations, with
a better and more tightly constrained fit to their ionization state,
allowing us to more precisely determine their abundances. Furthermore,
a strong Ca component, which was hinted at with one outlying point
above the \asca\  model, is firmly detected with the far superior
effective area of \xmm\ and \chandra . 
The only difference between the \asca\ and the \xmm\ and \chandra\ 
best-fit abundances lies in the
prediction for Ne, Mg and Fe. The \asca\  results for these species
were based solely on the 1-2 keV portion of the 
spectrum, where Ne and Mg K$\alpha$ lines and Fe-L lines should
lie. However, no strong evidence for particular lines was seen, and
the model preferred to fill in that region with Fe-L emission. The
\xmm\  and \chandra\ spectra, however, exhibit a prominent Mg line and 
should be sensitive enough to reveal an Fe-K line (at $\sim$6.7~keV), 
under the assumptions of
a single temperature planar model, if Fe were truly super-solar in
abundance.  Hence the deeper \xmm\ and \chandra\ spectra prefer the
combination of a less-absorbed, cooler continuum plus Ne and Mg lines
rather than a more highly absorbed continuum and Fe-L lines to fit the
soft emission. However, even this best-fit model only approximates, but does
not reproduce, the features around the He-like
Mg K-shell ``triplet'' at $\sim$1.34 keV.
For any single instrument, discrepancies from the model of similar magnitude 
occur across the entire spectrum (see the residuals in Figure
\ref{neifig}). However, for the most part the residuals from different
instruments do not all agree in sign or trend, whereas around 1.5~keV
and below all three exhibit a similar pattern (the MOS 2 spectrum is
similar to that of MOS 1). Combined with the fact that there are
two alternative minima in $\chi^{2}$ to explain the soft emission, it
is fair to say that the Ne, Mg and Fe results are less conclusive than
for Si, S, Ar and Ca. 

The new X-ray observations strongly indicate the presence of an
additional spectral component. The greater than 150 reduction in $\chi
^2$ for the addition of a powerlaw is highly significant (F-statistic
probability of chance occurence: 5.6$\times 10^{-20}$). 
Unfortunately the nature
of the second component is not constrained by these spectra. An
additional thermal component results in an equally viable fit (see
Section~7). 
For the purposes of this section the question is how has the addition
of a powerlaw component changed the fitted abundance pattern from the
\xmm\ and \chandra\ spectra.
As would be expected, the abundances increase systematically
when a second component is allowed to explain part of the continuum
emission. However the relative
abundances of Si,S, and Ar are virtually unchanged and [Ca/Si] is
increased but only from 3.09 to 3.4 [Ca/Si]$_{\sun}$. Without a second
component, the Ne and Mg abundances relative to Si drop below solar,
but their presence is still required for the fit. The Fe
abundance in the single component fit drops steeply partially 
because a higher electron temperature (1.1 versus 0.74 keV) is 
required to explain the high energy continuum.

Given our new X-ray parameters along with the more well-defined size
and distance from the radio observations we can refine our previous
estimates of the ambient density and age of the remnant to see if these
are consistent with a young ejecta-dominated SNR. 
For a first estimate, we assume that the normalization and temperature of
our single temperature fits are dominated by the blast-wave component
so that we can compare them to an analytical Sedov solution for the
expansion of a middle-aged SNR into a uniform density medium.
Following \citet{1998ApJ...505..732H},
we obtain the following values for the ambient density, $n_{\rm H}$, age,
explosion energy, $E_{0}$, and amount of swept-up interstellar material,
$M_{SU}$, given the electron temperature, $kT_{\rm e}$, normalization 
$N = n_{\rm e}n_{\rm H}V/4\pi D^{2}$, 
maximum angular radius $\Theta_{R}$, and an upper limit on the
distance to the remnant $D$: 
{\scriptsize
\begin{equation}
n_{\rm H} = 0.627
\left({{N}\over{2.0\times 10^{12}{\rm cm}^{-5}}}\right)^{1/2}
\left({{\Theta_{R}}\over{2.75\arcmin}}\right)^{-3/2}
\left({{D}\over{9.6 {\rm kpc}}}\right)^{-1/2}
{\rm cm}^{-3}
\end{equation}
\begin{equation}
age = 3653
\left({{kT_{\rm e}}\over{0.74 {\rm keV}}}\right)^{-1/2}
\left({{\Theta_{R}}\over{2.75\arcmin}}\right)
\left({{D}\over{9.6 {\rm kpc}}}\right)
{\rm years}
\end{equation}
\begin{equation}
E_{0} = 0.53 
\left({{N}\over{2.0\times 10^{12}{\rm cm}^{-5}}}\right)
\left({{kT_{\rm e}}\over{0.74 {\rm keV}}}\right)
\left({{\Theta_{R}}\over{2.75\arcmin}}\right)^{3/2}
\left({{D}\over{9.6 {\rm kpc}}}\right)^{5/2}
\times 10^{51}{\rm ergs}
\end{equation}
\begin{equation}
M_{SU} = 39.7
\left({{N}\over{2.0\times 10^{12}{\rm cm}^{-5}}}\right)^{1/2}
\left({{\Theta_{R}}\over{2.75\arcmin}}\right)^{3/2}
\left({{D}\over{9.6 {\rm kpc}}}\right)^{5/2}
M_{\sun}
\end{equation}
}
The upper limit on the distance implies that the remnant has reached the
Sedov stage, and that the spectrum should be dominated by the
outer-blast-wave moving into the ambient medium. On the other hand, if the
lower limit on the distance is used, the remnant is still quite young,
$\sim 750$ years, and the blast-wave would only have swept up $0.8
M_{\sun}$, but suggests an implausibly low explosion energy,
$E_{0}=1.0\times 10^{49}{\rm ergs}$ (even younger and less energetic
for the minimum angular radius).
There are two other potential indicators of the age of the remnant,
the abundances relative to solar and the ionization timescale. 
High-Z elements such as Ca have been shocked and yet the absolute
abundances are less than a factor of ten. The ionization timescale
measured is also longer than predicted by either end of the allowed
distance range. Both of these would seem to prefer older ages with a
greater amount of swept-up material. 
However they could also be explained by the reverse shock
having only reached a small portion of (well-mixed) ejecta, 
which if it were hot would bias the single temperature fit toward a
longer ionization timescale.  
Further discussion of 
the evolutionary state of \gtts\ and the interplay of ejecta
and ISM contributions to 
is deferred to
Section~7.1 when interpreting the detailed Type Ia
SNR models that explicitly incorporate the ejecta contribution.

\begin{figure*}
\noindent \includegraphics[width=6.4in]{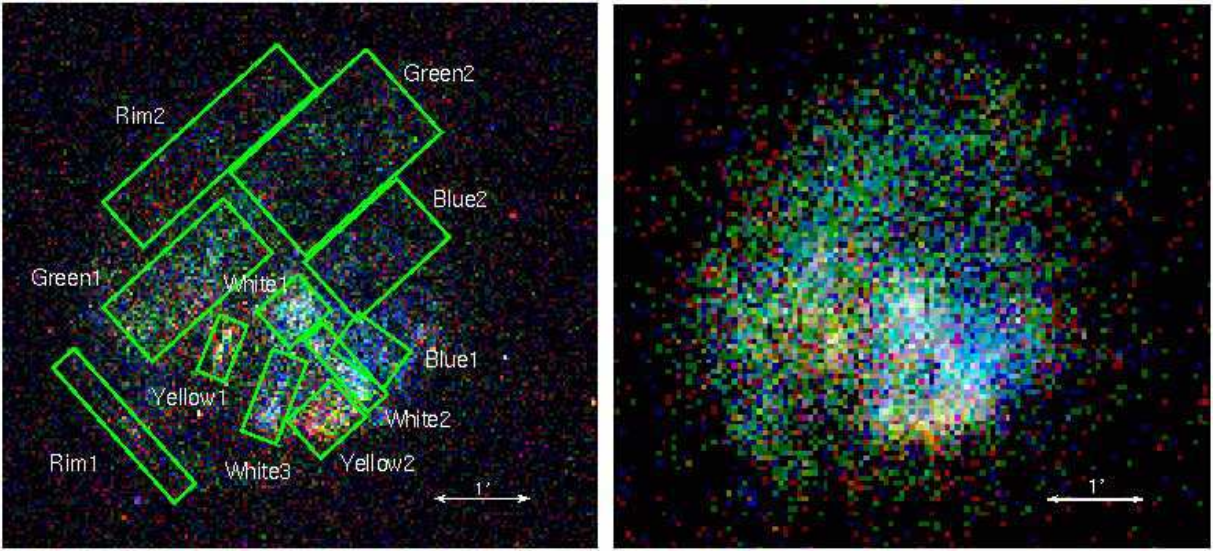}
\figcaption{Left: Three-color \chandra\ image of SNR \gtts: 
  red, 1.0 to 1.68 keV,
  green, 1.68 to 2.3, blue, 2.3 to 5.0 keV (binned by 4 pixels,
  linearly scaled from zero to half the maximum number of counts per
  bin in that band). These bands were chosen because they showed the
  largest variations from each other. The extraction regions chosen
  for spectral comparisons are as marked. \label{3colorbox}
  Right: A three-color image from \xmm\ MOS1 and MOS2, using the
  same three energy bands, but square-root scaling.
  \label{mos12_3color}}
\end{figure*}

Our NEI fits require that the ratio of Ca to Si is at least
3.4$^{+0.8}_{-0.75}$ times the solar value \citep{1989GeCoA..53..197A} 
at the 99\% confidence level. However, this is only true if our model 
assumptions hold; i.e. that the SNR can be adequately described by a 
single $kT_{\rm e}$, single final $n_{\rm e} t$, planar shock. 
Density and composition 
profiles in the ejecta, instabilities both in the explosion 
and in the SNR evolution, and profiles or inhomogeneities in the
ambient medium, should translate into different temperatures,
densities, and timescales as a function of position (and composition)
throughout the remnant. Such variations are suggested by the clumpy
morphology of \gtts\ in the X-ray band, 
an anchor-shaped bright region to the south
surrounded by smoother, more diffuse emission. To test the adequacy of
our single planar shock model, we first investigated how far
individual regions deviate from the globally averaged spectrum.

\section{X-ray Spectral Analysis: Spatial Variations within the 
         Remnant \label{vary}}

A comparison of images in small energy bands was used to identify
possible areas of spectral variation. Spectra from these regions were
first modeled independently of the global spectrum, each with a single
temperature, single timescale, variable abundance planar shock model. 
Even in these sub-sections of the remnant, the planar shock models
might not be representative of the real conditions in the ejecta, but
the uniformity of model assumptions allows for fair internal
comparisons between regions and to the global fitted parameters.
The individual variable-abundance fits are used to compare the
electron temperatures, ionization timescales and [Ca/Si] ratios
throughout the remnant independent of the global fit. 
Alternatively, we can test if plasma
conditions alone are sufficient to explain the variations by fixing
the relative abundances to the values found in the global fit and
allowing only the temperature, absorbing column density and ionization
timescale to vary. The aim of this search is twofold, first to identify
inhomogeneities in the SNR, but also to test if modeling the sum of
these disparate regions with a single temperature, timescale and
abundance set has biased our results.

\subsection{Extraction Regions}

Spectra extracted from the entire low-surface-brightness region 
show no obvious spectral differences from a spectrum containing all
the high-surface-brightness clumps, hence surface-brightness by itself
is not a good criteria for identifying spectrally distinct regions in \gtts .
Instead, narrow energy-band images from each instrument 
were examined separately  to
identify any spectral variations across the SNR.
Each event-file was divided into small spectral bins
covering individual lines and areas of continuum emission identified
in the overall SNR spectrum. Images in these bins were compared 
%using the three-color ``RGB'' facility of DS9 
to look for energy bands
whose images differed strongly from each other. Energy bands that
were indistinguishable spatially were combined for greater signal to
noise. The three energy bands that showed the greatest amount of
variability from each other are displayed in Figure \ref{3colorbox},
1.0 to 1.68 keV (red) 1.68 to 2.3 keV (green) and 2.3 to 5.0 keV (blue). 
This three-color image of \gtts\ separates the energy range
rich in Si and S lines (green) from the harder, more continuum
dominated, regime (blue) and
the soft band (red) which may include contributions from Ne, Mg, or Fe-L.
The largest fluctuations in color are within the bright regions, but similar
variability in the more diffuse region should not be ruled out, since
it might simply be missed at this level of signal to noise. The
differences seen within the bright region must be real features 
since they are repeated in both the \xmm\ and  \chandra\ observations.

\begin{figure*}
%\noindent \includegraphics[width=6.0in]{f6.eps}
\noindent \includegraphics[width=3.0in]{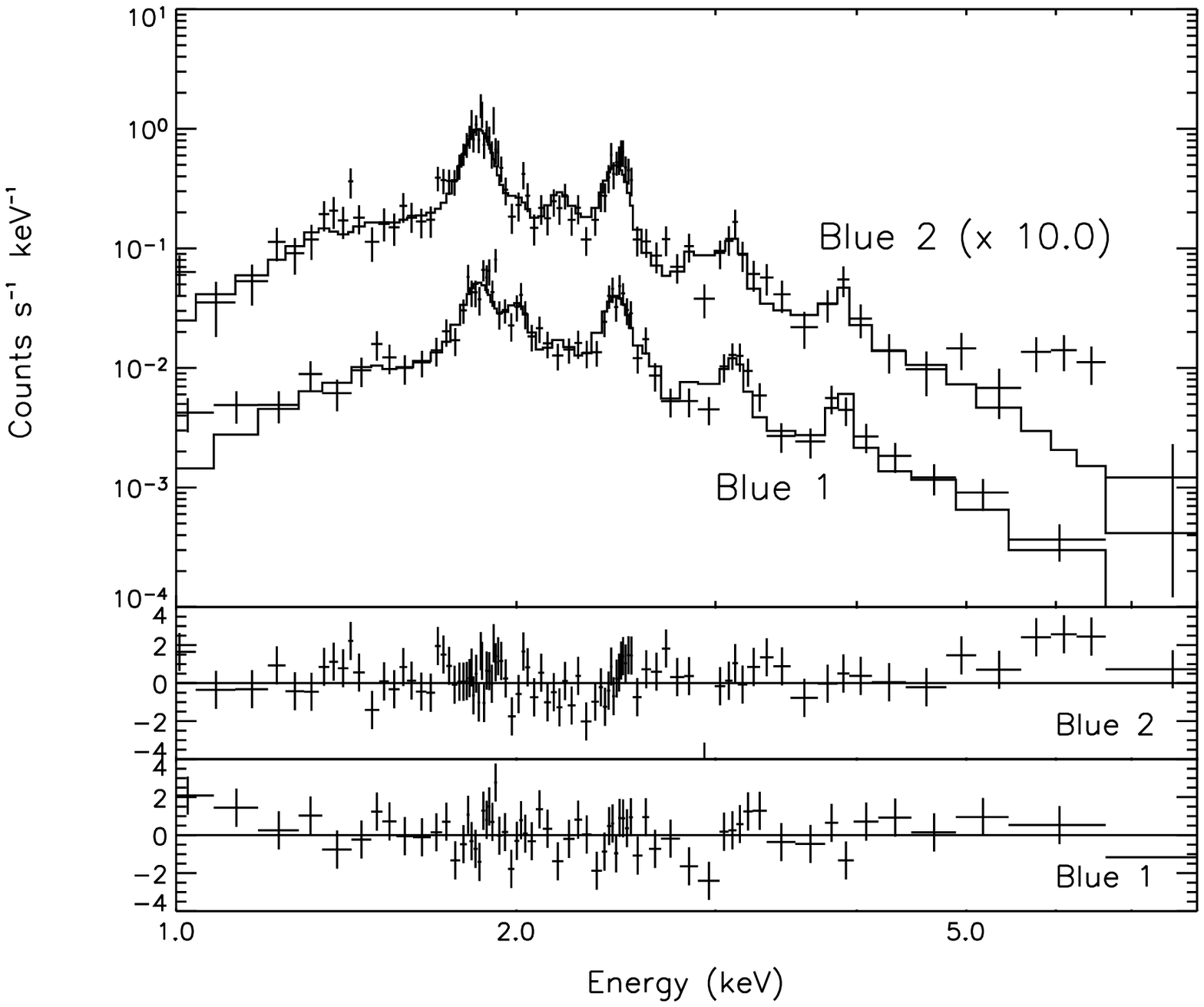} \hfill
\includegraphics[width=3.0in]{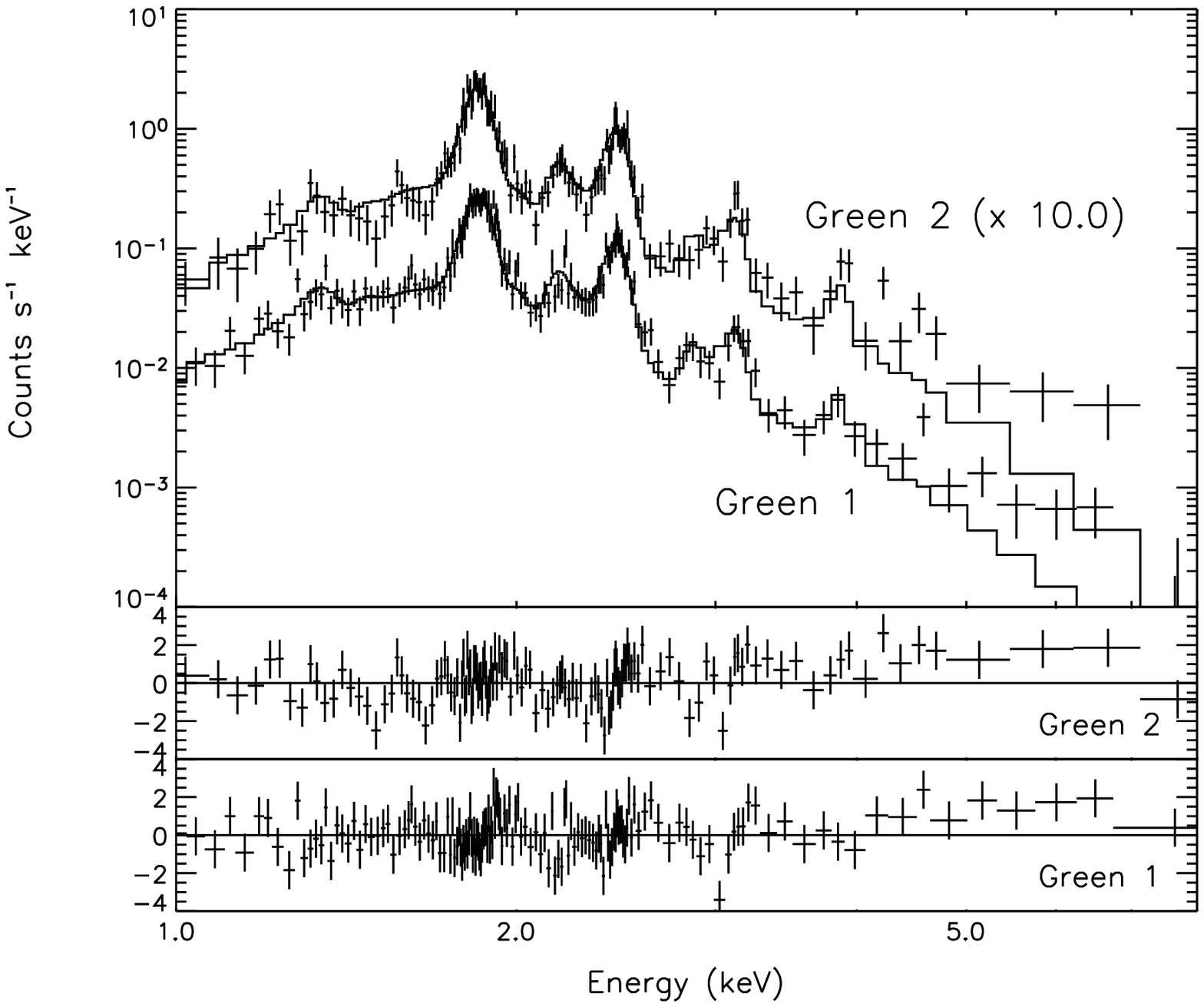} \\
\includegraphics[width=3.0in]{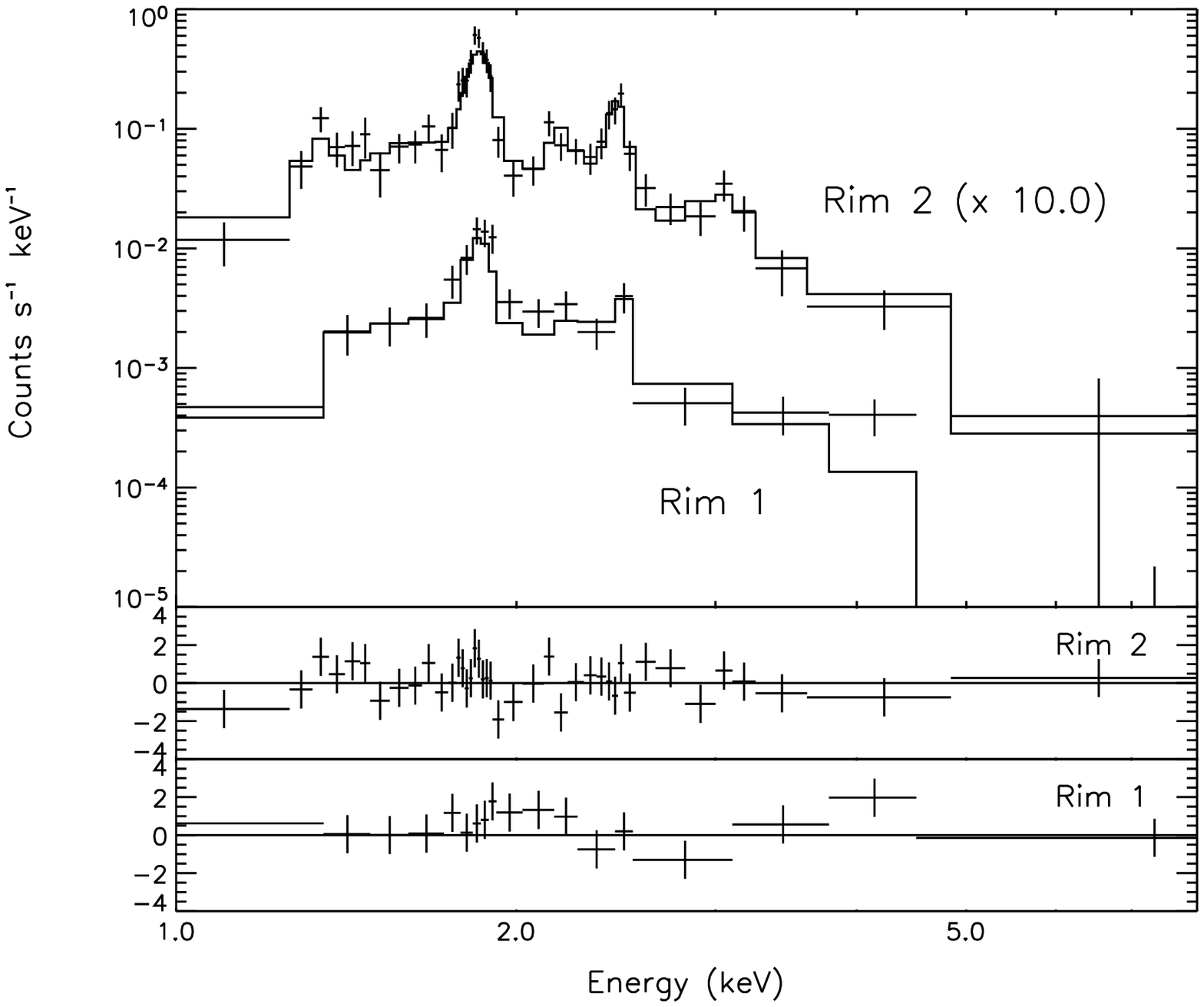} \hfill
\includegraphics[width=3.0in]{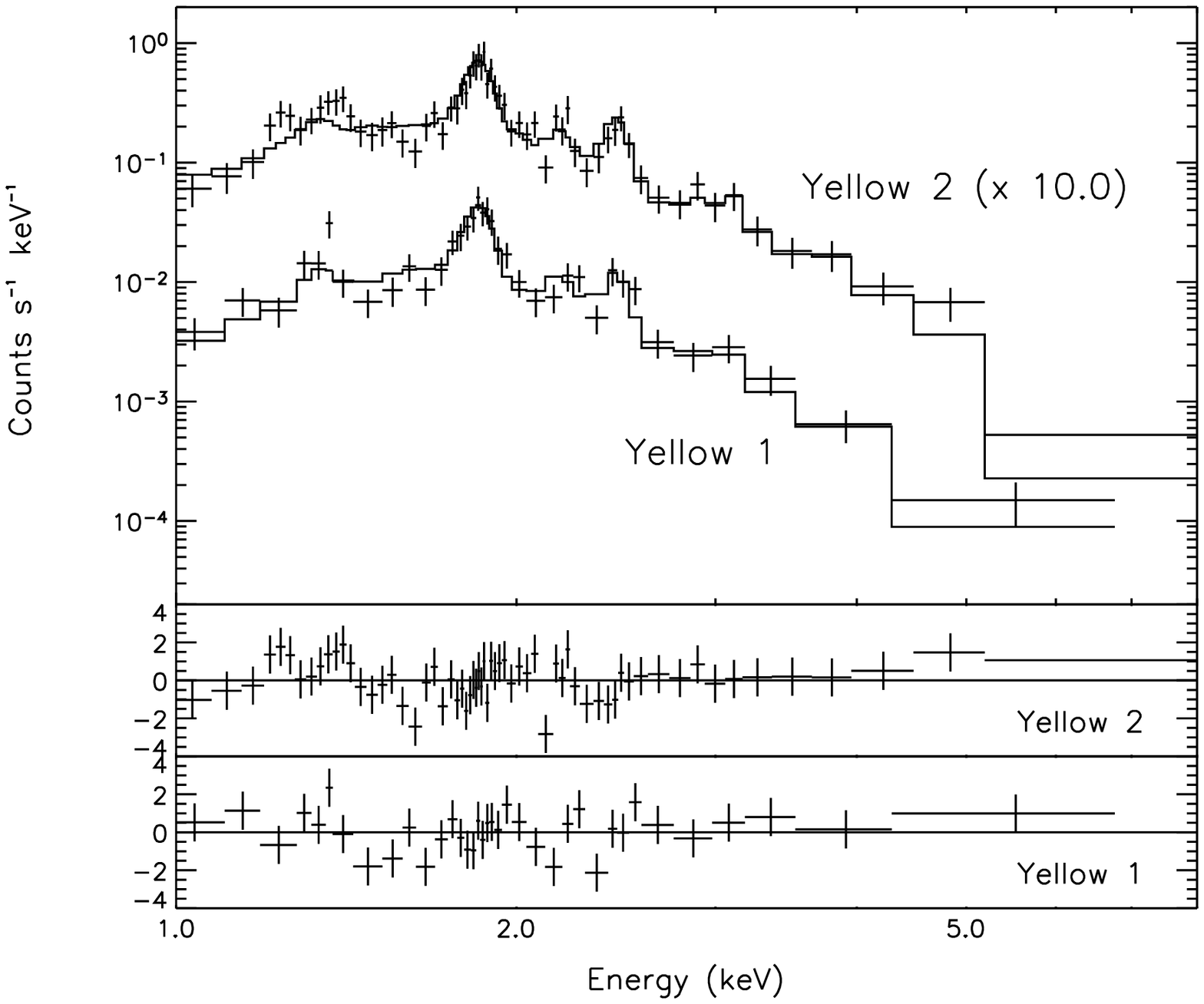} \\
\includegraphics[width=3.0in]{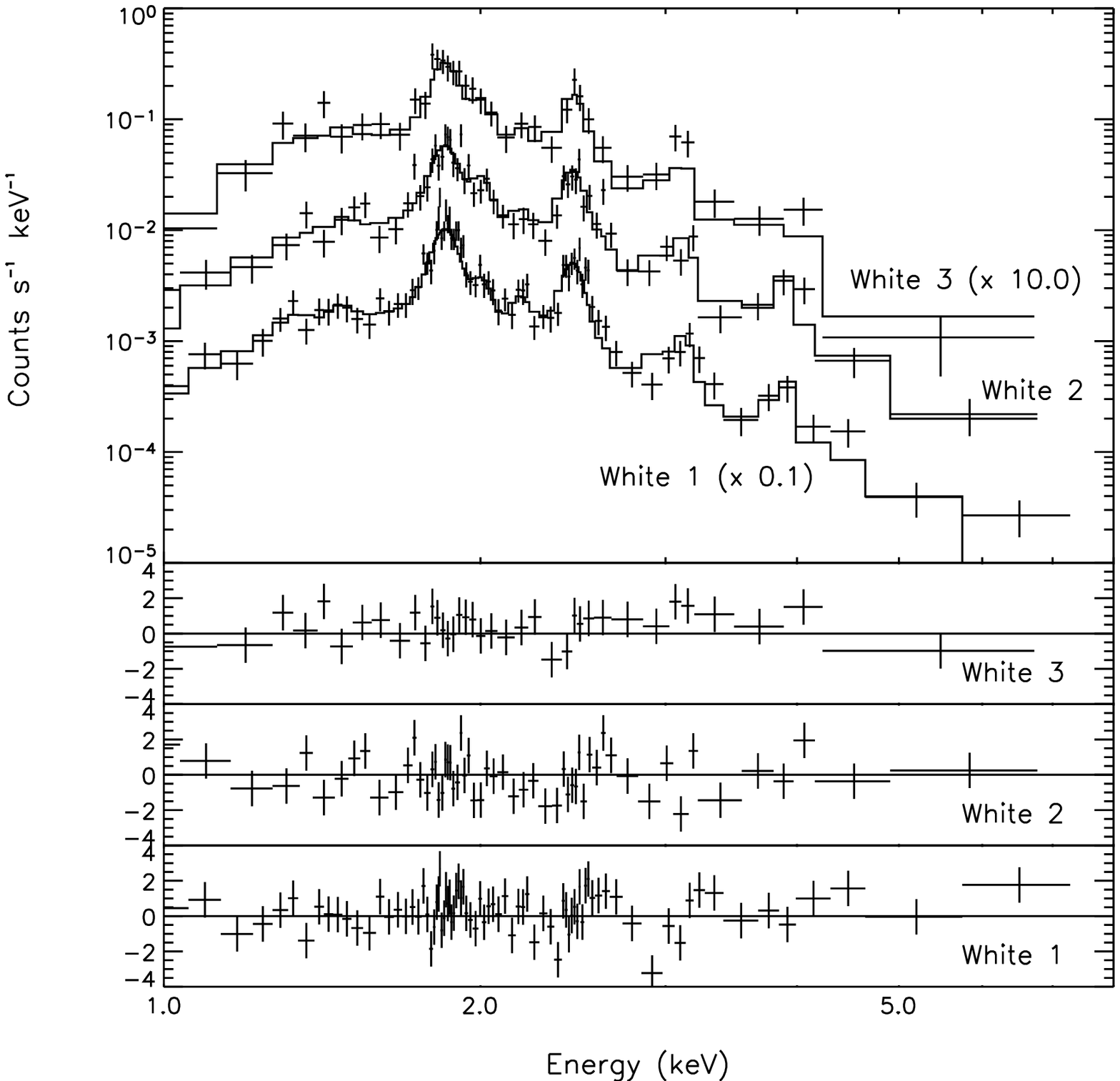} 
\figcaption{Spectra of eleven small regions around the
  SNR. In the NEI models shown here, the elemental abundance ratios
  were fixed to those found in the global SNR fit, without an
  additional power-law component. The spectra from all three \xmm\
  instruments and \chandra\ ACIS-S3  were used in constraining the
  best fit model, but only the \xmm\ EPIC PN spectra, which have the
  highest signal-to-noise, are plotted here for clarity. 
  \label{regionspec}}  
\end{figure*}

Eleven regions were chosen on the basis of their colors in Figure
\ref{3colorbox}:  
``blue1,'' ``blue2,'' ``green1,'' ``green2,''``rim1,'' ``rim2,''
``white1,'' ``white2,'' ``white3,''  ``yellow1,'' and ``yellow2.''
Spectra were extracted from 
these regions for each instrument, and individual response
matrices and ancillary response files were constructed using 
standard XMM-SAS and CIAO tools. 
Their spectra are shown in Figure \ref{regionspec}. All exhibit strong
Si and S lines indicative of enhanced abundances, but variations in 
conditions are evident in the
relative strength of the He-like line complex and the H-like line of
each species.  For instance ``green2'' shows little or no evidence of
any H-like Si, as opposed to ``white1'' where the 2.006~keV line from
H-like Si is relatively strong. 
Because the regions were chosen to isolate potential spectral differences
indicated by our image analysis, they do not all have similar numbers of
counts. Hence not all spectra can constrain the abundances equally well.

As before, the spectra were fit with a single component NEI model, with 
a variable electron temperature, ionization timescale and column density. 
For each region we first determined which elements required non-solar 
abundances as measured by the F-test for the addition of a new parameter. 
All regions showed significant signs of Si and S
enhancement, and many (blue1, blue2, green1, green2, white1, white2)
required an overabundance of Ca.  
No regions had statistics sufficient to
determine the Ne, Mg, Ar, or Fe abundances compared to their solar
values or constrain an additional power-law component. 
However, some do appear to have stronger Mg lines than others.

\subsection{Two Approaches and their Results}

In order to directly compare abundance measurements between regions, their 
spectra must be modeled under identical assumptions. 
Since the most
interesting results from the global fits were those of Si, S and Ca, we
chose to fit all regions with only those abundances free (all
parameters, Table \ref{regiontable}, selected values repeated in 
Table \ref{freevsSNR}). 
Overall, the fitted abundances of Si, S, and Ca are super-solar but 
tend to be slightly lower than in the global spectrum --- probably
because no power-law component was included in the small region fits. 
At 1$\sigma$ the fitted parameters indicate variation in the
electron temperatures ($0.61^{+0.11}_{-0.07}$ keV ``rim1'' to $1.17
\pm 0.04$ keV ``blue1''), logarithmic timescales ($11.09^{+0.17}_{-0.15}$
log(cm$^{-3}$s) ``rim2'' to fully equilibrated, $13.0_{-0.8}$
log(cm$^{-3}$s) ``white2''), and the [Ca/Si] abundance ratio ($1.9 \pm
0.5$ ``blue1'' to $4.0 \pm 0.8$ ``white2''). 
The detection of variations in the plasma parameters
holds at a much higher significance level than that of the abundances.
Also note that the fitted temperatures and [Ca/Si] ratios do not appear to be
correlated given that 7 out of 11 best-fit [Ca/Si] ratios lie between
3.69 and 4.22 over a wide range of temperatures.
From these fits it is clear that inhomogeneities in the conditions of
the emitting material exist.

\begin{deluxetable*}{lccccccc}
%\landscape
\tablecaption{Fits for small regions in \gtts\ \label{regiontable}}
\tablewidth{0pt}
\tabletypesize{\footnotesize}
\tablehead{
Region & $N_{H}$ & $kT_{\rm e}$ & log($n_{\rm e}t$)\tablenotemark{a}
 & Si & S & Ca & [Ca/Si] \\
 & ($10^{22}$ cm$^{-2}$) & (keV) & (log(cm$^{-3}$s) & & & }
\startdata
rim1\tablenotemark{b} 
& 3.65 (3.41--3.94) & 0.61 (0.54--0.72) & 11.67 (11.11--12.32)
& 1.50 (1.24--1.84) & 2.34 (1.93--3.03) & 11.76 (0--40.7) 
& 7.84 ($<30$)
\\
rim2\tablenotemark{b} 
& 3.35 (3.21--3.55) & 0.82 (0.72--0.87) & 11.09 (10.94--11.26) 
& 2.19 (1.95--2.40) & 3.27 (2.89--3.86) & 9.25 (1.94--16.93) 
& 4.22 (0.9--8.6)
\\
white1 & 3.01 (2.90--3.09) & 0.93 (0.91--0.95) & 12.0 (11.77--12.56) 
& 3.04 (2.80--3.32) & 4.13 (3.80--4.55) & 11.2 (9.02--13.58) 
& 3.69 (3.0--4.5)
\\ 
white2 & 3.36 (3.20--3.46) & 0.87 (0.85--1.02) & 13.0 (12.20--13.00) 
& 2.33 (2.13--2.68) & 3.61 (3.31--4.08) & 9.28 (7.34--11.30) 
& 3.98 (3.15--4.8)
\\
white3\tablenotemark{b} 
& 3.02 (2.92--3.17) & 0.90 (0.83--0.90) & 12.50 (12.17--13.00) 
& 2.10 (1.84--2.33) & 2.59 (2.29--2.87) & 5.86 (3.83--8.50) 
& 2.79 (1.8--4.2)
\\
blue1 & 3.84 (3.69--4.03) & 1.17 (1.13--1.21) & 12.13 (11.98--12.46) 
& 2.78 (2.52--3.08) & 3.48 (3.13--3.86) & 5.28 (3.91--6.65) 
& 1.90 (1.4--2.4)
\\
blue2 & 3.59 (3.37--3.76) & 1.08 (0.96--1.23) & 11.41 (11.25--11.57)
& 2.53 (2.28--2.89) & 3.24 (2.94--3.75 ) & 10.05 (7.6--13.1)
& 3.96 (2.9--5.2)
\\
yellow1\tablenotemark{b} 
& 3.37 (3.17--3.54) & 0.67 (0.61--0.76) & 11.48 (11.17--11.80) 
& 1.32 (1.17--1.50) & 1.31 (1.10 --1.55) & 5.40 (0--15.15) 
& 4.09 ($<12$)
\\
yellow2\tablenotemark{b} 
& 2.79 (2.69--2.89) & 0.87 (0.81--0.95) & 11.49 (11.36--11.64) 
& 1.50 (1.37--1.65) & 1.47 (1.30 --1.65) & 5.89 (2.61--9.38) 
& 3.93 (1.8--6.3)
\\
green1 & 3.24 (3.20--3.39) & 0.81 (0.74--0.85) & 11.51 (11.42--11.65) 
& 3.19 (2.93--3.39) & 4.60 (4.32--4.89) & 10.12 (7.76--12.69) 
& 3.17 (2.4--4.0)
\\
green2 & 3.87 (3.68--4.03) & 0.77 (0.72--0.85) & 11.49 (11.34--11.60)
& 3.63 (3.36--3.95) & 4.64 (4.25--5.11) & 14.60 (10.95--18.61)
& 4.02 (3.0--5.2)
\enddata
%\multicolumn{8}{l}{$^{a}$By $\log (n_{\rm e}t)=13.0$ ionization equilibrium
%  has been reached, hence no upper limit can be made on the age
%  or density for ``white2'' or ``white3.''} \\
%\multicolumn{8}{l}{$^{b}$the F statistic did not support the Ca 
%   abundance as an  additional free parameter in these regions}
\tablenotetext{a}{By $\log (n_{\rm e}t)=13.0$ ionization equilibrium
  has been reached, hence no upper limit can be made on the age
  or density for ``white2'' or ``white3.'' }
\tablenotetext{b}{the F statistic did not support the Ca 
   abundance as an  additional free parameter in these regions}
\end{deluxetable*}

The question at hand is whether the highly
super-solar [Ca/Si] ratio found in the best-fit to the global spectrum
could somehow be an artifact of naively summing spectra from regions with
varying plasma conditions.
The spectrum from region ``blue1'' indicates that there do exist areas
in the remnant where the [Ca/Si] ratio is consistent with the solar
ratio and inconsistent (at $1\sigma$) with the high [Ca/Si] ratio found
from the global spectrum. 
Other than this one region, the spectra are
consistent with the conclusions from the global model but are too
shallow to confirm it with any degree of certainty. At best, three regions,
``white1,'' ``white2'' and ``green2'' have [Ca/Si] ratios that exceed 3
times solar at the 1$\sigma$ level. Hence these fits have only shown
that high abundances of Ca relative to Si {\it may} exist throughout most of
the remnant, but have not substantiated the argument. 
On the other
hand, neither did we find a correlation between high [Ca/Si] ratios and
high temperatures or a single region especially rich
in Ca compared to Si that could account for the high Ca-K line flux in
the overall spectrum.

Alternatively we can model each of the individual regions with the
relative abundances found for the overall spectrum to see if any of
the spectra {\it require} abundances significantly different than the
overall fit or if all the variations can be fully accounted for by
changes in the conditions.
If we fix the relative abundances of Ne, Mg, Si, S, Ca, Ar, Fe, and Ni 
to the best-fit values for an NEI model of the global
spectrum (without a power-law component) but allow the overall
metallicity to be free, the spectra from all but one region 
(``yellow2'') are as well fit as they were under the previous set of
assumptions (Table \ref{freevsSNR}). 
Examining the spectrum of the discrepant region, ``yellow2,'' in
Figure \ref{regionspec} it appears that much of the disagreement with
the overall SNR ratios is at Ne and Mg. 
Using the F-statistic for the addition of 2 free parameters between 
the model which assumes the overall SNR abundance pattern and that
which allows Si, S, and Ca to vary freely, gives a 99.98\% probability that
the allowing for independent abundances of Si, S, and Ca is a true 
improvement for this region. 
Given that the fixed
abundance set used is not independent of the small region spectra, the
F-statistic is not strictly correct and actually underestimates the
significance of the reduction in $\chi ^{2}$. 

\begin{deluxetable*}{lcccccc}
%\rotate
\tablecaption{Small Regions modeled with overall abundance ratios
\label{freevsSNR}}
\tablewidth{0pt}
%\tabletypesize{\scriptsize}
\tablehead{ 
& \multicolumn{3}{c}{solar abundances} 
& \multicolumn{3}{c}{abundance ratios} 
\\
& \multicolumn{3}{c}{except variable Si,S,Ca\tablenotemark{a}} 
& \multicolumn{3}{c}{fixed at global SNR values} 
\\
region &  
$kT_{\rm e}$ (keV) & [Ca/Si]/[Ca/Si]$_{\sun}$\tablenotemark{b} & $\chi ^{2}$ (d.o.f.) & 
$kT_{\rm e}$ (keV) & Y/Y$_{SNR}$\tablenotemark{c} & $\chi ^{2}$ (d.o.f.) 
}
\startdata
rim1
& 0.61  & 7.84  & 121.11 (104)
& 0.84  & 0.65  & 123.61 (106)
\\
rim2
& 0.82  & 4.22  & 308.10 (271)
& 1.01  & 1.02  & 307.42 (273)
\\
white1 
& 0.93  & 3.69  & 305.18 (241)
& 1.14  & 1.29  & 309.59 (243)
\\ 
white2 
& 0.87  & 3.98   & 299.26 (212)
& 1.16  & 1.09   & 295.76 (214)
\\
white3
& 0.90  & 2.79  & 167.92 (173)
& 1.20  & 0.95  & 165.65 (175)
\\
blue1 
& 1.17  & 1.90  & 220.03 (200)
& 1.41  & 1.12  & 212.05 (202)
\\
blue2 
& 1.08  & 3.96  & 290.07 (275)
& 1.19  & 0.90  & 289.53 (277)
\\
yellow1
& 0.67  & 4.09  & 100.04 (116)
& 0.81  & 0.42  & 100.12 (118)
\\
yellow2
& 0.87  & 3.93  & 239.91 (205)\tablenotemark{d}
& 0.98  & 0.43  & 261.75 (207)
\\
green1 
& 0.81  & 3.17  & 530.43 (436)
& 0.92  & 1.55  & 531.24 (438) 
\\
green2 
& 0.77  & 4.02  & 405.68 (398)
& 0.844 & 1.72  & 400.01 (400)
\enddata
%\multicolumn{7}{l}{$^{a}$Selected values reproduced from Table 
%                         \ref{regiontable}. }\\ 
%\multicolumn{7}{l}{$^{b}$c.f. [Ca/Si]/[Ca/Si]$_{\sun}$ for SNR overall 
% without an additional power law component is 3.09.} \\
%\multicolumn{7}{l}{$^{c}$Ne, Mg, Si, S, Ar, Ca, and Fe were
% tied at their relative abundances from the global SNR fit. 
% Y/Y$_{SNR}$ represents a ratio of the form [Ne/H]/[Ne/H]$_{SNR}$ 
% for each of these species. }\\
%\multicolumn{7}{l}{$^{d}$The F-statistic for additional 
%   parameters from the fixed
%  abundance ratio model to the variable Si, S and Ca model is 9.33,
% probability of chance occurrence: 1.32$\times 10^{-4}$.}
\tablenotetext{a}{Selected values reproduced from Table 
                         \ref{regiontable}. }
\tablenotetext{b}{c.f. [Ca/Si]/[Ca/Si]$_{\sun}$ for SNR overall 
 without an additional power law component is 3.09. }
\tablenotetext{c}{Ne, Mg, Si, S, Ar, Ca, and Fe were
 tied at their relative abundances from the global SNR fit. 
 Y/Y$_{SNR}$ represents a ratio of the form [Ne/H]/[Ne/H]$_{SNR}$ 
 for each of these species. }
\tablenotetext{d}{The F-statistic for additional 
   parameters from the fixed
  abundance ratio model to the variable Si, S and Ca model is 9.33,
 probability of chance occurrence: 1.32$\times 10^{-4}$.}
\end{deluxetable*}

Given that the two prescriptions above identified two different 
regions as discrepant, we see that the conclusions of any given
analysis are effected by the choice of how to treat all the
abundances, {\it not just the abundances of  the elements with the most
  prominent lines.}  
The ``independent'' fits which allowed Si, S, and Ca to vary but held
all others at their solar values showed that the [Ca/Si] ratio for
``yellow2'' is entirely consistent with the overall SNR result despite
the poorer fit around the Ne and Mg lines 
when one requires {\it all} abundances to follow the SNR pattern (as
compared to the freely varying Si, S and Ca model).  
Vice versa, while the ``independent'' fits indicated that  ``blue1''
has a lower [Ca/Si] ratio than the overall SNR, this spectrum is
marginally better fit, $\triangle \chi ^{2} = 8.0$, assuming the overall
abundance pattern of the SNR rather than forcing some elements to be fixed to
their solar abundances.
In the case of ``yellow2'' the
abundances do not match the overall SNR, however it is not the high
[Ca/Si] ratio that is the problem. In the case of ``blue1'' the goodness
of fit to the overall SNR abundance ratio model shows that the initial
result of a lower [Ca/Si] ratio was an artifact of not properly
accounting for the emission from the less prominent species. However,
it is not just the abundances themselves that are affected by the
starting assumptions. The temperatures found using the global SNR
abundance ratios are systematically higher by about 0.2 keV than when
only Si, S, and Ca are freed from their solar values. This demonstrates the
magnitude of inherent bias that any given assumption about the
abundances can make in analyzing spectra with insufficient counts to
constrain all the relevant abundances.
The solution to this problem of a strong dependence on model
assumptions would be a full Bayesian analysis, but \gtts\ would not be
a good test case for such a technique.
For the present, the reader is urged
to treat abundance measurements with caution and to test the
sensitivity of their results to their underlying assumptions.

Despite variations in conditions, the average of the
metallicities of the small regions compared to the overall SNR model
is 1.01, or 1.14 when weighted by the total number of bins in each
spectrum. Hence, at least in this instance, using a single global
model has, if anything, underestimated the total abundances. Likewise
the Ca to Si ratio averaged over the small regions is 3.96 times the
solar ratio (or 3.77 weighted by the number of bins) which is also
greater than the value of 3.09 times solar found in the global SNR fit
before the inclusion of a power-law component.

\subsection{Implications for the Morphology of \gtts \label{csmring}}

Disregarding for the moment the low significance level of any
abundance variations, what would the fits to the small regions in
Tables \ref{regiontable} and \ref{freevsSNR} imply
about the X-ray and radio features discussed in Section 3?
The X-ray regions that most closely follow the bright radio ring in
both position and surface brightness are ``yellow1'' and
``yellow2.'' These two regions have the lowest abundances of Si, S,
and Ca found in the remnant and also the lowest metallicity in the
fits where the abundance ratios were fixed to the overall SNR values. 
Further, keeping Ne, Mg and Fe at their solar abundances produced a
significantly better fit for region ``yellow2'' than the sub-solar
overall [Mg/Si] values. Hence it is plausible to conclude that there is
more swept-up material in ``yellow1'' and ``yellow2'' than elsewhere 
and that the bright radio ring denotes an interaction with a denser
feature in the circumstellar or interstellar medium
(such as a circumstellar wind if the remnant is young or
an interaction with a molecular cloud if it is older) 
not the current
position of the reverse shock.  While the
abundances are higher for regions ``white1,'' ``white2,'' and ``white3,'' 
the nearly equilibrated ionization timescales support the
hypothesis of higher density material around the ring. 

The results from the two ``rim'' regions are less conclusive. While
``rim1'' does have the third lowest abundances, it still favors a high
Ca to Si ratio indicative of newly synthesized material. 
The larger region ``rim2'' shows no sign of 
lower abundances compared to the rest of the remnant, and the highest
absolute abundances are found in the unremarkable, diffuse emission 
regions ``green1'' and ``green2.'' 
This may be indicating that the reverse shocked ejecta and the
blast-wave into the ISM are no longer distinct in this remnant 
quite unlike other middle-aged SNRs such as DEM~L71.
On the other hand, if \gtts\ lies at the upper end of our distance
limits it would be far from the solar neighborhood, closer to the 
Galactic center where 
considerably more recycled material should enhance the metallicity 
by as much as 
0.3 dex \citep[e.g.][]{1983MNRAS.204...53S, 2000A&A...363..537R}.
The second solution may seem simpler, but it exacerbates the problem
of a super-solar Ca to Si ratio by requiring this abundance ratio to
have been obtained over the multiple SN explosions that produced the
metals in the current Galactic-center ISM.

\subsection{Summary}

In summary, the analysis of small regions throughout the remnant agree
with the conclusions of metal-enhanced material and a super-solar Ca
to Si ratio from the overall spectrum. While temperature and
ionization timescale differences are present, abundance variations are
comparatively small, as measured by planar shock models. 
In particular, we do not find any evidence for a
pocket of hot Ca that could explain the Ca-line flux in the overall
spectrum, nor do we see any strong deviations from the overall SNR
abundances except in a portion of the bright ring which appears to
have a greater contribution from swept up material.

\section{Potential SN Models \label{abund}}

Core-collapse and Type Ia explosions have widely different patterns of
abundances that should be readily detected in the X-ray emission from
their ejecta.
In core-collapse explosions the explosive nuclear burning front passes
through the existing hydrostatic layers of the star, burning them
incompletely and leaving large quantities of intermediate products like
O, Ne, Mg, Si, S, Ar and Ca. Much of the Fe-group material produced
may fall back onto the compact object since these are formed at the
highest temperatures and thus nearest the center, 
although some can be flung out
to large distances as seen in Cassiopeia A \citep{2000ApJ...528L.109H}. 
In Type Ia
explosions, on the other hand, the thermonuclear burning efficiently
incinerates most of the C/O white dwarf to Fe-group elements, with
fewer intermediate products and some unburnt C and O. There is no
compact object left behind in Type Ia's, so all the Fe-group material 
is ejected.  
Hence, traditionally the O/Fe ratio has been used to identify SNRs as
core-collapse or Type Ia explosions \citep[although the vastly
  different conditions of the O and Fe layers in a Type Ia can make
  this fitted ratio deceiving; see ][]{2003ApJ...593..358B}.
Here the question is whether a highly non-solar [Ca/Si] ratio could also be
diagnostic of the explosion mechanism.

In core-collapse SNe, Si, S, Ar and Ca are all primarily produced in
incomplete explosive oxygen burning, where their relative abundances 
are set to near solar values by quasi-equilibrium and do not vary 
significantly with conditions or mass \citep*{1973ApJS...26..231W}. 
Highly energetic or highly asymmetric explosions have larger
volumes of ``$\alpha$-rich freeze-out'' which produces $^{44}$Ti and
$^{44}$Ca. However, these explosions also have extended volumes of
explosive oxygen
burning such that the common Si and Ca isotopes dominate the [Ca/Si]
abundance ratio \citep[e.g. ][]{2002ApJ...565..405M,2003A&A...408..621K}.
For the prototypical core-collapse models of
\citet{1995ApJS..101..181W}\citepalias{1995ApJS..101..181W}, 
\citet{2002ApJ...576..323R}, and \citet{2003ApJ...592..404L}, 
the highest [Ca/Si] ratio found
($\sim$1.8[Ca/Si]$_{\sun}$) was for the 12 $M_{\sun}$ ``S12A'' model
of \citetalias{1995ApJS..101..181W}, hereafter WWS12A.\footnote{The
  total ejected mass ($1.32 M_{\sun}$) and kinetic energy of the
  explosion ($1.17 \times 10^{51}$ ergs) of WWS12A are quite similar
  to the values for Type Ia explosions, making a comparison on this
  basis largely moot.}  

For Type Ia models 
\citep[e.g.][]{1996ApJ...457..500H,1999ApJS..125..439I,2003ApJ...593..358B},
high [Ca/Si] ratios are found in delayed detonation explosions
(currently favored by the SN light-curves and spectra). 
In \citet{2003ApJ...593..358B}, the [Ca/Si] ratios of a delayed detonation,
DDTe,  and a pulsed delayed detonation,
PDDe, model are 2.5, and 2.9 times the solar ratio, respectively.  
Given the stratification of the ejecta present in the one-dimensional
models, the lack of observed Fe in \gtts\ can be explained by the fact
that Fe is shocked last and stays at a lower density 
\citep[see][for a discussion]{2003ApJ...593..358B}.
Recent 3-D deflagration models of Type Ia explosions
produce a similar
abundance pattern to the delayed detonation models shown here, without
an arbitrary transition from deflagration to detonation 
\citep[e.g.][]{2002A&A...391.1167R,2003Sci...299...77G,2005A&A...430..585G}.
However, they suffer from a shortage of intermediate-mass
elements \citep{2005AIPC..797..453B} and the burnt and unburnt
material are thoroughly bulk-mixed such that
it would be difficult to explain the lack of Fe emission in the data.

\begin{deluxetable}{lcccc}
\tablecaption{Relative Abundances: [X/Si]/[X/Si]$_{\sun}$ 
\label{totalejecta} }
\tablewidth{0pt}
\tablehead{
Ratio & \gtts\ & WWS12A\tablenotemark{a} & PDDe\tablenotemark{b} & DDTe\tablenotemark{b}}
\startdata
%O/Si &  & 0.17 & 0.03 & 0.07  \\
Ne/Si & 1.19 & 0.13 &   0.0010   & 0.015 \\
Mg/Si & 0.50 & 0.13 &   0.0017   & 0.025 \\
S/Si & 1.44 & 1.46  & 1.5 & 1.4 \\
Ar/Si & 0.49 & 2.20  & 0.68 & 0.60 \\
Ca/Si\tablenotemark{c} & 3.40  & 1.8  & 2.9 & 2.5 \\
Fe/Si\tablenotemark{d} & 0.28 & 0.37 & 0.89 & 0.91 \\ \hline
\enddata
\tablenotetext{a}{derived production factors after radioactive decay:
  \citet[table 6A]{1995ApJS..101..181W}.}  
\tablenotetext{b}{\citet{2003ApJ...593..358B}
Ne and Mg masses were not presented but are shown here.}
\tablenotetext{c}{2.65 - 4.2 at 99\% confidence.}
\tablenotetext{d}{0.14 - 0.46 at 99\% confidence.}

\end{deluxetable}

\begin{figure}
\noindent \includegraphics[width=3.2in]{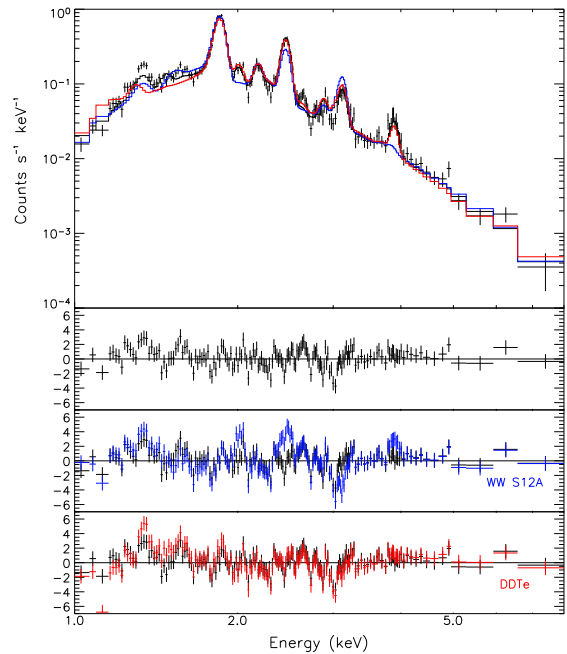} \\
\figcaption{Total ejected elemental masses compared to the global SNR
\gtts\ spectrum.   Only the MOS 1 spectrum is shown for simplicity,
although all four instruments were considered in the analysis. The
original free-abundances NEI model is shown in black. For comparison
the relative abundances were fixed to the appropriate ratios given the
total ejected mass of each element in a core-collapse (WW S12A: blue)
and a Type Ia (DDTe: red) model. 
\label{typeabund}}
\end{figure}

Table \ref{totalejecta} compares the fitted abundance ratios to the
total metal production from a core-collapse model, 
WWS12A and two Type Ia models, PDDe
and DDTe (recall the caveats to such a comparison from Section \ref{aside}). 
All models reproduce the S to Si ratio well. 
The core-collapse model matches the Fe to Si ratio well and does
produce some Ne and Mg, but strongly over-predicts the production of
Ar while still falling shy of the measured [Ca/Si] ratio. 
The [Ar/Si] and [Ca/Si] ratios are especially problematic
because high production factors of Ca isotopes were accompanied by
high production 
factors of Ar in all of the solar abundance massive star models 
of \citetalias{1995ApJS..101..181W}.    
Both Type Ia models clearly match the Ca to Si ratio best and do not
conflict as badly concerning the production of Ar. On the other
hand, Fe is strongly over-predicted and Ne and Mg are strongly
under-predicted by the Type Ia models. To determine the degree to
which these discrepancies can be seen in the spectrum we  performed
fits of the global spectrum with the abundance ratios of C, N, O, Ne,
Mg, S, Ar, Ca, Fe and Ni (relative to Si) fixed to the SN
explosion predictions. Figure \ref{typeabund} compares NEI models for
the spectrum that use the abundance ratios relative to Si 
of WWS12A or a delayed detonation (DDTe) model by
\citet{2003ApJ...593..358B}.   
The deviations from the data for WWS12A are both more prominent and
harder to explain away.
While either model produces a good fit to the data if Ar (WWS12A) or Fe
(DDTe) are allowed to vary, the stratification of Type Ia models
provides a natural motivation for reducing the amount of Fe seen,
whereas Ar, Ca, and Si should be primarily co-spatial in the ejecta
of a core-collapse explosion.  
Furthermore, swept-up solar composition material will add Ne and Mg
needed by both models but should not flip the sense of the Ar to Si ratio
for WWS12A. 
Tests of a two-component model, one with solar composition (ISM) the other
with freely varying abundances (ejecta) bear this out. 
If the temperatures of the two components are set to reasonable values
(such as the blast-wave and EM-averaged Si temperatures
from Section~7.1) the fitted [Ne/Si], [Mg/Si] and [Fe/Si]
abundance ratios in the ``ejecta'' component drop steeply, while
the sub-solar [Ar/Si] ratio  and super-solar [Ca/Si] and [S/Si] ratios
remain at values similar to the one component fit. Hence a strict
comparison of the global abundance ratios clearly favors an origin as
a Type Ia SN. 
If true, \gtts\ would be only the third potential Ia SNR with a
significantly disturbed morphology after Kepler's SNR
\citep[e.g.][]{2004A&A...414..545C} and N103B
\citep{1995ApJ...444L..81H,2003ApJ...582..770L}, both of which also
have controversial identifications.

A core-collapse scenario with solar-like abundance ratios between Si,
S, Ar, and Ca could still be viable if the strong Ca-K line is the
result of differing plasma conditions rather than abundance. Line
fluxes in general are strongly dependent on the electron temperature and
ionization, hence if the temperature of the Ca-rich material were
sufficiently higher than in the Si-rich ejecta this could also account
for their unusual line ratio. 
A global fit allowing a separate temperature and ionization
time-scale for Ca but holding [Ca/Si]=[Ca/Si]$_{\sun}$ performs
equally well as our single temperature fit with a free Ca
abundance, with almost identical best-fit $\chi^{2}$ values. 
The electron temperature associated with the Ca emission was \cate ~keV as
compared to \site ~keV for the Si and other elements.
However, no spatially distinct region with a high temperature and
high [Ca/Si] ratio was identified in the small region fits. On the other
hand, this does not preclude the possibility of two unresolved or
co-existing plasmas. 
For instance, one could consider the case of a two density medium in
pressure balance, with a single elemental composition shocked at a
single time. We looked at a model with two NEI
components with the same abundances, their temperatures tied at a
series of fixed ratios and the ionization timescales offset
appropriately for pressure balance. This
did result in a sub-solar [Ne/Si] ratio ($\sim$0.8[Ne/Si]$_{\sun}$) and
reduced the amount of Fe required by the fit
($\sim$0.05[Fe/Si]$_{\sun}$) but did not change the pattern of Si,
S, Ar and Ca abundances substantially ($\sim$1.5[S/Si]$_{\sun}$, 
$\sim$0.5[Ar/Si]$_{\sun}$, $\sim$3.0[Ca/Si]$_{\sun}$). 
Hence, although this question is not well-defined, it seems unlikely
that a multi-temperature plasma 
with an abundance pattern typical of a core-collapse explosion
could explain the global spectrum of \gtts\ without considerable 
compositional differences from place to place.

In the above discussion 
we have compared the global spectrum of \gtts\
to the total metal output of the SN explosion models
with the addition of a few simple two-temperatures tests and concluded
that the products of a Type Ia explosion are more consistent with the
spectrum. 
In reality the blast-wave into the ISM may contribute substantially to
the total spectrum and the 
ejecta itself will not be a homogeneous mix of the nucleosynthesis
products. Layering of the composition or large scale anisotropies imply
that not all products will be visible in the X-rays at all times, nor
will they be under uniform conditions. A true comparison of the 
global SNR spectrum to explosion models requires that the
evolution of ejecta and shock fronts be followed from the time of the
explosion until the present age of the remnant to predict the X-ray 
spectrum. While no such model is available for the variety of core-collapse 
models, the models of \citet{2003ApJ...593..358B} provide the opportunity for
exactly such a comparison for the case of a Type Ia explosion.

\subsection{Investigation of Possible Type Ia Explosion 
Mechanisms \label{Iasec}}

Many groups have produced grids of Type Ia models to compare to SN
spectra and lightcurves
\citep[e.g.][]{1996ApJ...457..500H,1999ApJS..125..439I}.
\citet{2003ApJ...593..358B} have expanded upon this work to enable
comparisons to the remnants of Type Ia SNe.
\citet{2003ApJ...593..358B} followed the plasma conditions and
ionization state within the ejecta of Type Ia SNe from the explosion
itself to 5000 years later, and calculated the expected X-ray emission
of the ejecta and the blast-wave for various times in the evolution of
the remnant. 
Their network of
Type~Ia explosion models attempted to cover all currently debated
explosion mechanisms from pure detonation (super-sonic flame
propagation) to pure deflagration (sub-sonic flame propagation), as
well as a sub-Chandrasekhar mass explosion. 
Their models showed a strong evolution of the contribution
from each element to the X-ray spectrum. Indeed, for some models Fe
did not begin to appear until late times.

The structure of the ejecta as prescribed by each SN Ia model will
evolve as the forward shock expands out into the ambient medium and a
reverse shock is generated that propagates back into the layers of
ejecta. The X-ray emission from the SNR will have contributions from
both the outer blast wave (reasonably approximated by a Sedov solution
after on the order of 1000 years) 
and the reverse shock \citep[followed in detail in
  the models of][]{2003ApJ...593..358B}.  
In this simple one dimensional picture only the outermost
layers will be shocked at the earliest times, and thus begin to ionize
and emit in the X-rays. Elements that are only present in deeper
layers will show up later in the spectrum and experience different
temperatures and densities than the earlier layers. 

For comparison with the spectrum of \gtts ,
for every explosion model in the grid of \citet{2003ApJ...593..358B}, the
evolution of the SNR was computed in a grid of three ambient
densities, $\rho _{am}$
(0.2, 1.0, 5.0 $\times 10^{-24}$ g cm$^{-3}$). From each of  these
SNR models, spectra from the shocked ejecta were extracted at four
different ages (500, 1000, 2000, and 5000 years). 
The synthetic spectra of 
Badenes et al. are completely characterized by four parameters only: the 
Type Ia explosion model, the age of the SNR $t$, the AM density 
$\rho_{am}$, and the amount of collisionless heating at the reverse 
shock $\beta$ (defined as the ratio of electron to ion specific internal 
energies at the shock front). Among these parameters, we have not 
explored variations in the last one ($\beta$), which tend to enhance the 
Fe emission, and particularly the flux in the Fe K blend, due to the 
higher electron temperatures brought on by collisionless heating. In the 
case of \gtts , this effect is disfavored by the conspicuous absence of Fe 
K in the X-ray spectrum. For a discussion on collisionless electron 
heating at SNR shock fronts, see \citet{2005AdSpR..35.1017R}, 
for its impact on the 
ejecta emission, see sections 2.2 and 2.4 in \citet{2005ApJ...624..198B}.

One significant advantage of the spectra of
\citet{2003ApJ...593..358B} is that they
have an underlying physical model for the dynamics of the SNR. However,
the underlying model is calculated using approximations that need to be
taken into account for the comparison between the data and the synthetic
spectra. One such approximation is the fact that the models are
calculated in one dimension. 
This is clearly a simplification of a very complex
situation, because deviations from spherical symmetry are expected both
on theoretical grounds (the contact discontinuity between the forward
and reverse shocks is subject to Rayleigh-Taylor instabilities) and due to
the large scale structure observed in the X-ray images of \gtts\ 
(see sections 3 and \ref{csmring})
Thus, the constraints obtained by the comparison between the spatially
integrated spectrum and the one dimensional models of 
\citet{2003ApJ...593..358B}  are focused on
the emission measure averaged emission of the SNR (i.e., they tend to
reflect the properties of the brightest regions). Local deviations in
the properties of the ejecta and blast wave emission from the best-fit 
synthetic spectrum are expected, but the average properties of the
plasma should be similar to those of the most successful models. Since
we consider the emission from all the elements at the same time, it is
unlikely that this kind of local deviations would conspire to affect
our choice of models, but this possibility cannot be ruled out
completely. Another important approximation in the models is that the
ambient medium  is assumed to have a uniform density. Again, this is
just a simplification of a complex scenario. Theoretical studies
indicate that the circumstellar medium around a Type Ia SN should be
strongly modified by the pre-supernova evolution of its binary
progenitor, which would sculpt a bubble-like cavity
\citep{2001ApJ...556L..41B}.  This should affect the dynamics of the
subsequent SNR \citep{2000ApJ...541..418D}, but the observations of
prototype Type Ia SNRs like Tycho or SN1006 show no evidence for such
strongly modified dynamics 
\citep[][, see also Badenes 2004, PhD thesis]{2001ApJ...556L..41B}. On
the other hand, the fact that Type Ia SNe are not detected in the
radio \citep{1989ApJ...336..421W}, 
or X-rays \citep{1993ApJ...412L..29S}
seems to indicate that the
impact of the progenitor evolution on the circumstellar medium is
limited. 
Only two Type Ia SNe have shown signs of H$\alpha$ emission from
interaction with circumstellar material 2002ic
\citep{2003Natur.424..651H} and 2005gj \citep{2005IAUC.8633....1P}. 
In the absence
of strong evidence to the contrary, we consider the constant ambient density
hypothesis as a valid approximation, and we note that if this were not
the case, our detailed comparisons might need to be revised. 
For reference, given emission-measure and metallicity considerations, 
the density of the total ambient medium swept-up in the ring is about 
twice that of the diffuse emission \citep[see][for discussions of how
  the mass and radius of a wind shell might effect the dynamics of the
  SNR]{2005ApJ...630..892D}.

The spectra were
calculated using a plasma emission  code maintained by K. Borkowski,
whose inputs are the electron temperature  and ionization structure of
each ejecta layer as predicted by the SNR model 
\citep[see][for details]{2003ApJ...593..358B}.   
An additional Sedov component
appropriate for the specified age, ambient density and explosion
energy was used to model the blast-wave contribution. 
The version of the emission code used here 
is missing calculations for Argon as well as some of the Li-like 
lines for other elements that can be important for material far from
ionization equilibrium. (These were included in the NEI code used in the
previous section.)  In particular we believe the disagreement in the line
centroid for Ca in the upcoming comparisons is an artifact of the
atomic data used in this NEI emission code. 
However, for the purpose of comparing the models
with the observed global spectra, over a discrete grid in density and age,
these emission models were the most straightforward to use and readily
available to the community. Only the
normalization and absorbing column density are allowed to vary to find
the best fit for each grid point.

The $\chi ^{2}$ statistic is inappropriate for these comparisons
since we are not allowing the parameters to vary freely.
These models have a more limited range of flexibility to fit the
observations than the general NEI fits used before
given that the evolution and abundance pattern are
entirely constrained.
However, we can
still use the best-fit  $\chi ^{2}$ value at each grid point as a quantitative
guide for which models are a closer match. It is also useful to record
the $\chi ^{2}$ value over energy bands within the spectrum to confirm
our suspicions based on visual inspection of why a certain age and
density is ``fitting'' well or poorly. These numbers are not of
rigorous statistical value, but rather just a rough measure of which
models are good matches and why.

Just like the fits presented in Section~5, 
the goal is to match the abundances,
ionization state and temperature of the observed spectrum. However in
this case, with two connected but separately evolving components,
matching the data is a matter of finding the best balance of these
components for any given model. For a given density, as time
progresses the Sedov component becomes more dominant, reducing the
apparent equivalent widths of the metal lines, and the blast-wave
component also cools, eventually to below the temperature we measured
in the last section. Meanwhile the ejecta component is also
evolving. The reverse shock is penetrating deeper into the layers of
nucleosynthetic products, and the already-shocked material is
continuing to equilibrate in temperature and ionize. 

\begin{figure}
\noindent \includegraphics[width=3.2in]{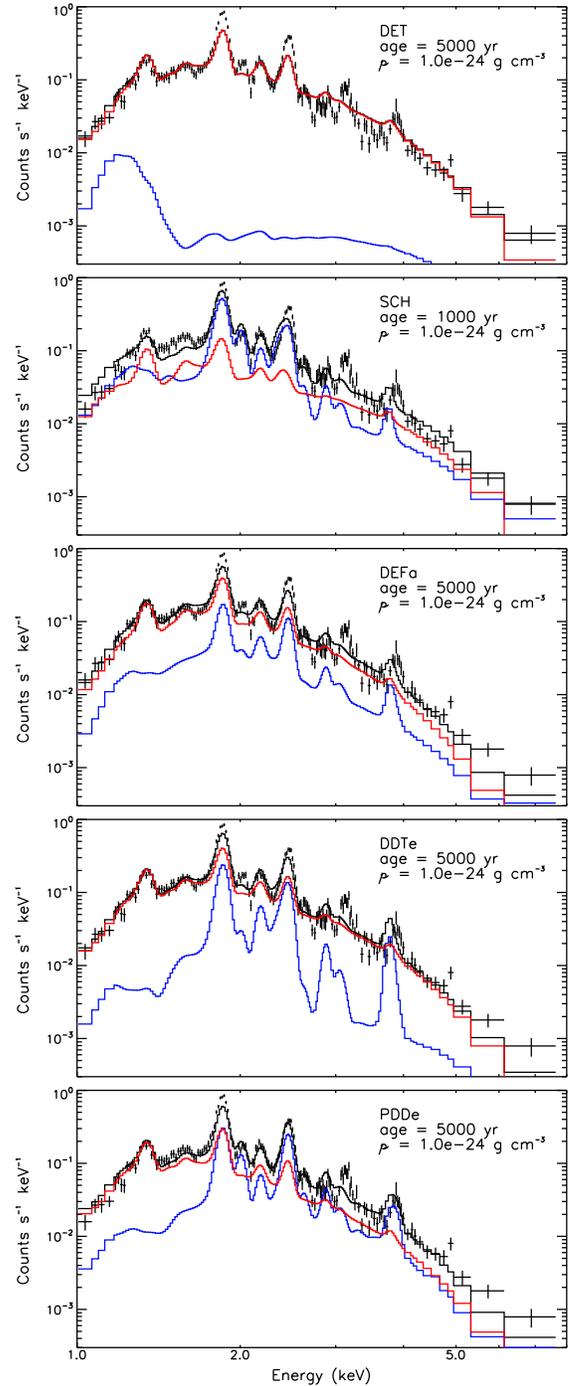} 
\figcaption{Type Ia SNR model comparisons over the original grid of
  densities and ages. The blast-wave (red) and ejecta (blue) both
  contribute to the overall model spectrum (black). The closest
  matches for each explosion type are plotted here against just the 
  MOS1 data for clarity. 
From top to bottom,
DET pure detonation, $\rho_{am}= 1.0\times 10^{-24}$ g cm$^{-3}$, 
age = 5000 years; 
SCH sub-Chandrasekhar, $\rho_{am}= 1.0 \times 10^{-24}$ g cm$^{-3}$, 
age = 1000 years.
DEFa pure deflagration, $\rho_{am}= 1.0 \times 10^{-24}$ g cm$^{-3}$, 
age = 5000 years.
DDTe, delayed detonation, $\rho_{am}= 1.0\times 10^{-24}$ g cm$^{-3}$, 
age = 5000 years and 
PDDe, pulsed delayed detonation, $\rho_{am}= 1.0\times 10^{-24}$ g cm$^{-3}$, 
age = 5000 years.
The bottom two panels show the delayed detonation models that 
are a better match to the continuum shape, line fluxes and ionization
state than the others.
\label{Iafailed}}
\end{figure}

The first test a model must pass is whether there is a time when the
abundance pattern of the ejecta (modified by the solar
abundance contribution for the blast wave) matches that inferred from
the data and the relevant emission lines are of appropriate strength
relative to the continuum (from both
the Sedov and ejecta components). Three explosion models can be
rejected from this consideration alone (see Table~\ref{SNIatable}). 
A pure detonation explosion
(DET) efficiently burns all material to Fe-group elements as it
propagates supersonically through the white dwarf, leaving only a
thin skin of intermediate elements before it is quenched in the
outer layers. Thus for DET the iron produced always dominates
the Si, S, and Ca present in the spectrum, entirely unlike the lack of
iron (relative to most Ia models) found in \gtts . The closest match
occurs once the solar abundance ambient medium is dominating (Figure
\ref{Iafailed}a) which we already know cannot produce strong enough
lines to reproduce the spectrum.

In the sub-Chandrasekhar model (SCH) a detonation originates on the
surface of the star and burns the outer layers of accreted
material. When the ensuing pressure waves converge on the center of
the WD, a second ignition happens and a detonation propagates back
out. Both the
center and the outside of the star are incinerated to Fe-group
elements, while the middle region is less completely burned. 
For SCH, there
simply isn't a time at which the abundance pattern is a good match for
\gtts . Prior to $\sim$1000 years the Sedov component is insufficient to
fill in the soft continuum. Around $\sim$1000 years the Si lines are
reasonably well matched in ionization state and strength relative to
the continuum (see Figure \ref{Iafailed}b) but no other portion of the
spectrum is well-modeled at this time. After $\sim$1000 years the Si and S
ejecta are swamped by the Sedov component. This blast-wave dominated
nature occurs earlier for the SCH models than it does for the other
Type Ia models because of the low ejected mass and explosion energy.

The pure deflagration models (DEF) are also a poor match to the spectrum of
\gtts . Here the flame front is sub-sonic, and the
the star expands in reaction to the flame propagating out from the
center. The flame is quenched in the outer regions when the expansion
of the material is of the same order as the flame velocity. This front
burns the interior efficiently to Fe-group elements but leaves a large
portion of unburnt C and O in the outer layers.
The high column density to \gtts\  masks the emission lines of
carbon and oxygen so that the preponderance of this unburnt material
in the pure deflagration models manifests itself as strong continuum
emission from 1 to 8 keV. Thus even within the ejecta component the
equivalent widths of Si, S, and Ca are fairly low. 
The Si-rich material lies deep enough in the star that by the time 
Si emission is prominent the Sedov contribution is starting to
dominate (Figure \ref{Iafailed}c). Furthermore since it has been
shocked relatively recently the ionization state of Si is not
sufficiently advanced.  Soon after the DEF
``best match'' the Sedov contribution dominates and the Fe-K emission
becomes too prominent as well. Simulations with a finer grid in
ambient density and age for the SCH and DEF models did not improve 
upon the fits shown in Figure \ref{Iafailed}.

The remaining viable models are the delayed detonation and pulsed
delayed detonation models, DDT and PDD, that were preferred initially
because of the likeness of their overall abundance pattern to that of
\gtts . In delayed detonation models, the flame begins at the center as a
deflagration, briefly forcing the expansion of the star, but at some
prescribed flame density it transitions to a detonation front. 
Like the DEF models the center is efficiently burned
to Fe-group elements, but due to the initial expansion the detonation
burning in the outer layers encounters a wider region where intermediate mass
elements can be produced before the flame is quenched. 
Thus there is a shell of Si-rich material outside the Fe-rich core. 
The PDD models invoke a pulsation of the star to induce the transition
from deflagration to detonation. They produce a similar pattern of
elemental production to DDT with a modified density structure.
In both cases a range of transition densities were explored, with the
models labeled from a to e in order of the earliest to the latest
transitions from deflagration to detonation. 

\begin{figure}
\noindent \includegraphics[width=3.4in]{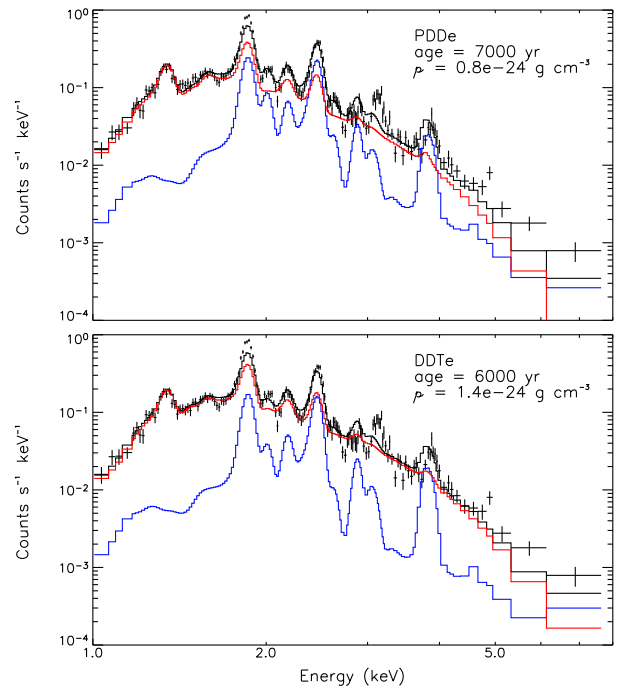} 
\figcaption{The ``best match'' models for DDTe and PDDe over a finer
grid of densities and ages. As in Figure \ref{Iafailed}, the
blast-wave (red) and ejecta (blue) both contribute to the overall
spectrum (black). For DDTe $\rho_{am}= 1.4 \times
10^{-24}$ g cm$^{-3}$, age = 6000 year. For PDDe $\rho_{am}= 0.8 \times
10^{-24}$ g cm$^{-3}$, age = 7000 year. Only the MOS 1 spectrum, which
has the highest spectral resolution around the Si line-complex, is
shown, although all four instruments were considered in the
analysis. \label{bestmatch}} 
\end{figure}

As seen in the previous section, the primary disagreement
between the overall elemental production and the inferred abundances in
\gtts\ was the over-prediction of Fe. 
Since the Fe is produced
primarily in the more interior portion of the star in either delayed
detonation model, it is possible that the computed emission from
Badenes' models might more accurately match the \gtts\ spectrum than
the naive comparison to the total elemental production. 
As expected, amongst the delayed detonation models, PDDe and DDTe that
produce the least amount of Fe, are preferred
over their more energetic cousins when comparing their predicted
spectra with that of \gtts . 
In fact both PDDe and DDTe have fairly good matches to the \gtts\
spectra at an ambient density of $1.0 \times 10^{24}$ g cm$^{-3}$ and
an age of 5000 years 
(Table \ref{SNIatable}, Figure \ref{Iafailed}d and e).  
In comparison to the DET, SCH, and DEF models that have ``failed'',
the delayed detonation models produce stronger Si, S, and Ca lines at
the appropriate ionization stage than any of the others at close to
the right ``abundance'' ratios. The S lines are quite well matched and
Si is under-predicted but less so than in DEF or DET. 
At this stage, the
reverse shock has actually penetrated fully through the ejecta, but
the Fe remains ``hidden'' due to a lower electron temperature and ionization 
timescale (see below for details) and the dominance of the
blast-wave component. 
%Table \ref{SNIakTnt}

\begin{deluxetable*}{lcccccccc}
%\rotate
\tablecaption{Best ``matches'' for each Type Ia model \label{SNIatable}}
\tablewidth{0pt}
\tabletypesize{\scriptsize}
\tablehead{
model  & $\rho_{am}$ & age & $r_{fwd}$\tablenotemark{a} 
& $D_{r}$\tablenotemark{b} & 
$N_{\rm H}$ & norm & $D_{norm}$\tablenotemark{c} & $\chi^{2}$ \\
 & $10^{-24}$ g cm$^{-3}$ & years & pc & kpc 
& $10^{22}$ cm$^{-2}$ & (at 10kpc) & kpc & 
(just MOS2)\tablenotemark{d} }
\startdata
DET & 1.0 & 5000 & 12.2 & 15.5--18.2 & 3.96 & 0.73 & 11.7 & 1247.31 \\
DEFa & 1.0 & 5000 & 9.8 & 12.5--14.6 & 4.12 & 1.99  & 7.1 & 786.44 \\
SCH & 1.0 & 1000 & 5.7 & 7.3--8.5 & 3.17 & 1.4 & 8.4 & 795.15 \\
PDDe & 1.0 & 5000 & 11.7 & 14.9--17.5 & 3.22 & 0.44 & 15.1 & 672.26 \\
DDTe & 1.0 & 5000 & 11.5 & 14.6--17.2 & 3.78 & 0.77 & 11.4 & 701.68 \\ \hline
PDDe & 0.8 & 7000 & 14.0 & 17.8--20.9 & 3.75 & 0.94 & 10.3 & 549.33 \\
DDTe & 1.4 & 6000 & 11.5  & 14.6--17.2 & 3.80 & 0.53 & 13.7 & 627.66 
\enddata 

%\multicolumn{9}{l}{$^{a}$the forward shock radius implied by Badenes's model
%at that age and density }\\
%\multicolumn{9}{l}{$^{b}$the distance to the SNR derived by comparing
%$r_{fwd}$ with the angular radius of the radio remnant (2.7\arcmin\ to
%  2.3\arcmin ) }\\
%\multicolumn{9}{l}{$^{c}$the distance to the SNR derived from the
%normalization, $D_{norm} = (norm)^{-1/2}$ 10 kpc }\\
%\multicolumn{9}{l}{$^{d}$Given only as a guide, cannot be interpreted in the
%  usual sense }
%\caption{
%$^{a}$the forward shock radius implied by Badenes's model
%at that age and density \\
%$^{b}$the distance to the SNR derived by comparing
%$r_{fwd}$ with the angular radius of the radio remnant (2.7\arcmin\ to
%  2.3\arcmin ) \\
%$^{c}$the distance to the SNR derived from the
%normalization, $D_{norm} = (norm)^{-1/2}$ 10 kpc \\
%$^{d}$Given only as a guide, cannot be interpreted in the
%  usual sense 
%}
\tablenotetext{a}{the forward shock radius implied by Badenes's model
at that age and density }
\tablenotetext{b}{the distance to the SNR derived by comparing
$r_{fwd}$ with the angular radius of the radio remnant (2.7\arcmin\ to
  2.3\arcmin ) }
\tablenotetext{c}{the distance to the SNR derived from the
normalization, $D_{norm} = (norm)^{-1/2}$ 10 kpc }
\tablenotetext{d}{Given only as a guide, cannot be interpreted in the
  usual sense }
\end{deluxetable*}

The next question is whether there is an age and density where the
apparent ionization state of Si, S, and Ca are reached before the
Sedov component swamps the ejecta component. To evaluate this
requires examining a more densely spaced grid in ambient density and
age. For PDDe, the $\rho _{am} = 1.0 \times 10^{24}$ g cm$^{-3}$
model, at 5000 years appeared to be very close to the correct
ionization timescale, but possibly under-abundant in Si, S and Ca. 
Hence we looked to slightly lower densities
where the Sedov solution would take longer to become dominant, but
towards longer times so that the model would reach the same ionization
stage ($\rho _{am} = 0.8, 0.9, 1.0 \times 10^{24}$ g cm$^{-3}$ , age =
5000, 6000, 7000, 8000 years).  For DDTe at an ambient density of $1.0
\times 10^{24}$ g cm$^{-3}$ and age of 5000 years, the ionization
state of Si and S both appear too young. There is no evidence for a
H-like Si component above the Sedov continuum. Hence we looked to
slightly higher densities and ages around 5000 years for a better
match ($\rho _{am} = 1.2, 1.4, 1.6, 1.8 \times 10^{24}$ g cm$^{-3}$,
age = 4000, 5000, 6000, 7000 years).  For all these models, a
post-facto test of the importance of cooling shows that neglecting
radiation is a valid approximation for these densities and ages. 

Using the $\chi ^{2}$ value relative to all three \xmm\ spectra (MOS1,
MOS2 and PN) as a guide we find a best match for PDDe at $\rho_{am} =
0.8 \times 10^{24}$ g cm$^{-3}$, age = 7000 years ($\chi ^{2}$
= 2287) and for DDTe at $\rho_{am} = 1.4 \times 10^{24}$ g cm$^{-3}$,
age = 6000 years ($\chi ^{2}$ = 2526) as shown in Figure
\ref{bestmatch} 
(the implications of these particular parameters will be discussed shortly).
Given that the only variables in these models are the
ambient density, the age and the absorbing column density, these are
remarkably good matches for the character of the \gtts\ spectrum.
While the Si abundance or flux in the He-like Si K-shell lines 
($\sim$1.86~keV) are less than they should be, the H-like Si
Ly$\alpha$(2.0~keV) and Ly$\beta$(2.37~keV) are well matched by the
combination of the blast-wave and ejecta components. 
S and Ca are primarily contributed from the ejecta component. Both 
compare well with the data in flux, and S appears to agree in ionization
state as well. Ne, Mg and the soft continuum are dominated and
well-modeled by the blast-wave. The hard end of the continuum is
slightly under-predicted but no strong Fe-K line from the ejecta is
found that would disagree with the \gtts\ spectrum. Comparing to the
previous figure, the failures in the other models may become
clearer. 
Once the first requirement was met, i.e. the lack of a prominent Fe-K
line, the most diagnostic aspects of the spectrum were the
soft-continuum, H-like Si-K, and the S and Ca abundances. 

The relative contributions from the blast wave and ejecta to 
individual spectral features are also interesting to note in light of our
previous discussions on the interpretation of the global and small
region spectra.  In particular the best-matched delayed detonation
models attribute essentially all of the Ne and Mg line emission to the
solar-composition blast wave component, exactly in line with our
explanation for their abundances in the global fits. 
This is also in
keeping with the interpretation of the bright ring as a circumstellar
interaction given that the Ne and Mg emission from region "yellow2"
preferred solar abundances as opposed to the sub-solar abundances
relative to Si seen in the remnant as a whole.

\begin{deluxetable}{lcccc}
\tablecaption{Electron Temperatures and Ionization Timescales \\ 
   for Two of the Type Ia Models \label{SNIakTnt}}
\tablewidth{0pt}
%\tabletypesize{\scriptsize}
\tablehead{
 &  Sedov\tablenotemark{a}  &  \multicolumn{3}{c}{Ejecta\tablenotemark{b}}  \\
 &         & Si & Ca & Fe 
}
\startdata
\multicolumn{5}{l}{PDDe \tablenotemark{c}}\\
$kT_{\rm e}$ (keV) 
     & 0.69  & 1.68  & 1.83 & 1.60 \\
log($n_{\rm e}t$) log(cm$^{-3}$s) 
     & 11.58 & 11.12 & 10.63 & 10.26 \\
\multicolumn{5}{l}{DDTe \tablenotemark{d}}\\
$kT_{\rm e}$ (keV) 
     & 0.65  & 2.11  & 2.11 & 1.82  \\
log($n_{\rm e}t$) log(cm$^{-3}$s) 
     & 11.76 & 10.53 & 10.53 & 10.27 \\
\enddata
\tablenotetext{a}{Mean shock temperature and ionization age}
\tablenotetext{b}{Volume emission measure weighted averages for each element}
\tablenotetext{c}{$\rho _{am}$: 0.8$\times 10^{-24}$ g cm$^{-3}$  
                   age: 7000~years}
\tablenotetext{d}{$\rho _{am}$: 1.4$\times 10^{-24}$ g cm$^{-3}$  
                   age: 6000~years}
\end{deluxetable}

Given that the line fluxes, continuum shape and apparent abundances
here are the sum of multiple components, how do the electron
temperatures and ionization timescales compare with our naive single
NEI model?  
Table~\ref{SNIakTnt} lists the temperatures and ionization timescales
for the Sedov blast-wave component and selected elements at the
specific age and ambient density that best-match the spectrum of
\gtts . The parameters listed for the ejecta are the EM
weighted averages of the electron temperatures and ionization
timescales in the regions containing those elements.
Given the evolutionary state implied by the ambient density and ages
found here it is not surprising that fits to the global spectrum
reflect the blast-wave more than the ejecta.
The temperatures and timescales we measured under the
assumption of a single $kT_{\rm e}$ and $n_{\rm e}t$ most closely approximate 
the Sedov component of the Type Ia SNR models, but may have been
pushed towards higher values of $kT_{\rm e}$ and $n_{\rm e}t$  by the bright
hot ejecta.  %
Naively one might worry that the stratified abundance pattern
that minimizes the Fe-K line flux would also imply that the Ca should
be under significantly different plasma conditions than Si. 
However, the EM averaged
temperatures and ionization timescales for Ca and Si are actually
quite similar to each other in PDDe, and almost identical in DDTe
because these elements are largely co-spatial in the ejecta (see
Table~\ref{SNIakTnt}). 
On the other hand, the EM averaged quantities for Fe
are quite different from Si, especially given that both the lower electron
temperature and the shorter ionization timescale imply a lower level
of ionization. These differences combined with the
relative dominance of the blast-wave component allows the Fe to remain
hidden. 
None of the ionization timescales in these Type Ia SNR models reach
the value from the single temperature NEI model, however the Sedov
ionization timescales are quite similar to the values found in the
fainter small regions outside the bright ring.

Badenes's Type Ia SNR models naturally provide
two determinations of the distance to the SNR by predicting both the
radius of the outer-blast-wave, which can be compared to the angular
size of the radio SNR, and the EM normalized to a nominal
distance of 10~kpc. 
Comparing the two distance estimates for a single Type Ia model 
(Table \ref{SNIatable}) we see that they 
are at least grossly comparable with each other
(not surprising given that the angular size used is from the faint 
material, while the normalization is largely due to the bright
material). The distances estimated here for the PDDe and DDTe models 
from the normalization and radius are larger than but similar to 
the upper limit on the distance from the 
\ion{H}{1} absorption measurements. 
Also note that the ages and densities of the best-matched Type Ia
models are roughly the same as those estimated in Section~\ref{OneNEI}
for the far end of the allowed distance range. However,
unlike the estimates in Section~\ref{OneNEI} which
only considered emission from the blast-wave, here the ejecta
component is explicitly considered and a pre-defined explosion energy
has been set. In fact these ages and ambient densities were primarily
chosen to match the relative importance of Sedov and ejecta components
seen in the spectrum. 
Given a reasonable distance estimate of 10~kpc 
the radius of the bright ring would then be 4.6~pc. 
While this is considerably larger than 
a typical planetary nebula, it is actually quite similar to the
size of some of the largest planetary nebula with
measured parallaxes \citep{1990ApJ...360..173B}, 
consistent with the interpretation of the ring as an interaction with
a structure in the circumstellar medium.

Simply increasing the ambient density to increase the
ionization timescale and decrease the radius will not solve the
remaining discrepancy in distance because an increased density also 
results in an increased luminosity which in turn implies a larger
distance to the SNR \citep[explained in detail by][]{2005astro.ph.11140B}.
However, the history of the remnant is not likely to be as
simple as we have modeled. Radiative losses to the shock, energy lost
from the escape of cosmic-rays, the blast-wave running into a 
dense circumstellar ring of material or ambient cloud, would all be
consistent with a decreased inferred shock energy, a higher timescale
and a smaller radius. It is likely that more than one of these has
occurred given that roughly three quarters of the explosion energy
would need to be dissipated to explain the discrepancy between the
Type Ia models and the \ion{H}{1} distance range. 
If on the other hand general absorption in the region around \gtts\
has obscured a true absorption feature at $-116$\kms\ then no
modification is required and we are left with a best estimate of the
distance from the X-ray spectrum and radio extent alone.

\section{Conclusions \label{discuss}}

The line-rich thermal \asca\ X-ray spectrum of SNR \gtts\
identified it as an ``ejecta-dominated'' remnant 
\citep{2001ApJ...548..258R}. In this paper we examine the new \xmm ,
\chandra , and \atca\ observations of \gtts\ to learn as much as
possible about its progenitor system.

The radio observations provide an image of the full extent of the remnant.
At both 1.4 and 5~GHz the remnant exhibits diffuse emission
throughout a shell 2.3\arcmin $\times$ 2.7\arcmin\ in radius that is
limb-brightened in some places but not in others. In the interior lies a
bright elliptical ring of radio emission that must indicate a
local increase in either density or magnetic field (or both). 
Compared to the X-rays, we see at least one outer-rim radio
filament with X-ray emission, and there is bright X-ray emission
related to the radio ring but brightest in different places. Radio
\ion{H}{1} absorption measurements also provide a constraint on
the distance of between $2.0\pm 0.5$ kpc and $9.3\pm 0.3$ kpc.

The X-ray spectrum of SNR \gtts\ has been studied in detail both globally
and locally. Starting from a single temperature NEI planar shock model
for the spectrum of the whole remnant, 
the previously seen super-solar abundances of Si and S are confirmed
and further constrained (\sibf\ and \sbf\ times their solar abundances,
respectively). The most striking new features of \xmm\ and
\chandra\ spectra are a strong Ca K-shell line, and the lack of any
significant K-shell line from Fe.  
The lack of an Fe-K line implies that the soft X-ray
spectrum is coming from a cooler bremsstrahlung continuum plus a
Mg-K line, rather than unresolved Fe-L lines as previously argued by
\citet{2001ApJ...548..258R}. The strong Ca line 
indicates the presence of Ca-rich ejecta, yielding a Ca abundance of
almost 12 times the solar value in the single temperature, single
timescale model. Moreover the global spectrum requires a [Ca/Si] 
ratio of at least \casill\ times solar at the 99\%
confidence level. In contrast, the [Ar/Si] ratio is found to be sub-solar.

We looked for but could not find any evidence to doubt the abundance
pattern of Si, S, Ar, and Ca found in the single temperature
model. Spectra from small regions around the remnant showed variations
in electron temperature and ionization timescale but no clear evidence for
variations in the abundance pattern. 
The abundances in the bright radio ring are if anything lower than the
surrounding areas indicating an origin in the circumstellar medium. 
For the global spectrum two
additional simple two-temperature models were explored, one with
abundances tied across components and one that had a solar
composition plasma in addition to the freely varying one. These
differed from the single temperature model in the relative abundances
of Ne, Mg or Fe but not of Si, S, Ar, and Ca.

The abundance pattern seen in the global spectrum 
strongly favors a Type Ia origin. The composition and
stratification of the ejecta in delayed-detonation Type Ia models can
explain the overall abundance pattern if the contribution from a
solar composition blast-wave into the ISM is included and the reverse
shock has only recently reached the bulk of the Fe-group ejecta. 
On the other hand, typical core-collapse
explosions produce Si-group elements in the same nuclear burning
processes and at
approximately their solar ratios, inconsistent with the abundances of
our single temperature global SNR model. Core-collapse explosion
models that do have a highly super-solar [Ca/Si] also have 
super-solar [Ar/Si] ratios, the opposite of which is seen in \gtts . 
The only remaining viable explanation
for a core-collapse model is if a significant amount of hot Ca-rich
material is segregated from the bulk of the Si-rich plasma. Our
examination of small regions throughout the remnant cannot exclude
this possibility especially if the two plasmas were unresolved 
but no spatially distinct hot region with 
a significantly high [Ca/Si] ratio was found.

As emphasized above, 
a single temperature model is inherently overly simplistic
in that it does not reflect the variations in composition and conditions
within a SNR either between the blast-wave and the reverse-shocked
ejecta or within the ejecta itself. Given that our initial results support 
an origin in a Type Ia SNe and that \citet{2003ApJ...593..358B} have
produced predicted spectra from Type Ia SNRs, we have the
opportunity to test whether our conclusions hold when actual
conditions are taken into account. 
Simulated spectra from a variety of Type Ia explosion models  
were compared with the global spectrum of \gtts\ to find the ambient
density and age for each model that best matches the spectrum, 
i.e. that avoids a strong Fe-K line, reproduces the strength
of the He-like Si-K lines relative to the continuum, exhibits bright
S and Ca lines, and reflects the line shapes and
line-ratios within a given element.
In the end we
saw that indeed Type Ia models that did not initially match the
fitted abundances in total metal production also did not match the
spectrum at any special time in their evolution. The preferred delayed
detonation and pulsed delayed detonation models could in fact reproduce
the generic features of the \gtts\ spectrum. In these simulated
spectra the reverse shock has over-run the Fe-rich material, but there
is still no strong Fe-K line because Fe is at a lower
EM-weighted average temperature than the other elements and
the high energy X-ray band is dominated by blast-wave continuum
emission. 

From the above analyses combined with distance constraints 
from the radio data, we are also able to infer the current
evolutionary state of \gtts . 
Simple Sedov estimates from the single temperature, single timescale
fits yield ages of between 750 and 3500 years.
For the Type Ia SNR models of \citet{2003ApJ...593..358B},
based on the relative contributions of
the interstellar material and the ejecta,
\gtts\ is about 5000 years old. Both the age and normalization of the Type
Ia SNR models imply that \gtts\ is probably toward the upper limit of
its allowed distance range.
Hence, while \gtts\ may not be ``ejecta-dominated'' in a literal sense, even 
after 5000 years and with more than 50~$M_{\sun}$ of swept up 
material, \gtts\ still exhibits clear evidence of ejecta.

To reiterate the main purpose of the paper, analysis of the X-ray
spectrum of \gtts\ have allowed us to identify it as a Type Ia remnant
and, if our 
identification is correct, show that low-energy (Fe-poor) delayed 
detonations or pulsating delayed detonations are the preferred explosion 
mechanisms. The conspicuous absence of Fe K emission in the X-ray 
spectrum is a strong indication that little Fe is found in the outermost 
layers of SN ejecta, where it would be revealed by the high densities 
and electron temperatures. This precludes any Type Ia SN model with 
plumes of Fe ($^{56}$Ni) at high velocities, including sub-Chandrasekhar 
explosions and three dimensional deflagrations with well-mixed ejecta. 
We also find that the bright ring of radio emission is likely 
to be evidence of the impact of the Type Ia progenitor system on the 
circumstellar medium. In the case of \gtts , the high [Ca/Si] ratio,
combined with a low [Ar/Si] ratio  
found in the single ionization, single electron temperature models, did 
indeed point to a Type Ia origin - a scenario which then was shown to be 
viable under a more detailed scrutiny.
However, the low abundance of Fe, which one
might have expected to have come from a core-collapse explosion, was
actually a red-herring, and could be explained in the Type Ia
scenarios once actual conditions within the ejecta were taken into
account.

\acknowledgments

The authors would like to thank Naomi McClure-Griffiths for her
assistance in the use of the Southern Galactic Plane Survey data. 
This research has made use of SAOImage DS9, developed by the
Smithsonian Astrophysical Observatory 
\citep[\url{http://hea-www.harvard.edu/RD/ds9/}]{2003ASPC..295..489J} 
and the XSPEC spectral fitting package 
\citep[\url{http://xspec.gsfc.nasa.gov/}]{1996ASPC..101...17A}.
 C.E.R. was supported during this work by NASA grant NAG5-9281.
This work was also made possible by \chandra\ GO grant G02-3070X 
and NASA \xmm\ grant NAG5-9990.
{\it Facilities:} \textit{ASCA}, 
  \textit{Chandra X-ray Observatory (ACIS-S3)}, \textit{XMM-Newton
  (MOS1, MOS2, PN)}, \textit{ATCA}. The Australia Telescope is funded
  by the Commonwealth of Australia for operation as a National
  Facility managed by CSIRO.

%% Included in this acknowledgments section are examples of the
%% AASTeX hypertext markup commands. Use \url without the optional [HREF]
%% argument when you want to print the url directly in the text. Otherwise,
%% use either \url or \anchor, with the HREF as the first argument and the
%% text to be printed in the second.
%
%#\url{http://www.aas.org/publications/aastex} or the
%#\anchor{ftp://www.aas.org/pubs/}{AAS ftp site}.



\bibliographystyle{apj}   
\bibliography{bib_SNR_SNe}

%  \typeout{******************************************}
%  \typeout{** Please run "bibtex \jobname" to optain}
%  \typeout{** the bibliography and then re-run LaTeX}
%  \typeout{** twice to fix the references!}
%  \typeout{******************************************}

\clearpage

\end{document}